\documentclass[10pt,letterpaper]{amsart} 

\usepackage{color}
\usepackage{multicol}
\usepackage[letterpaper, margin=1in]{geometry}
\usepackage{subfig}
\usepackage[overload]{empheq}

\usepackage{xcolor}

\usepackage{listings}
\usepackage{amsmath, amsthm, amssymb, wasysym, verbatim, bbm, graphics}
\usepackage{enumerate}
\usepackage{url}
\usepackage{mathtools}
\usepackage[colorlinks,linkcolor=blue]{hyperref}
\usepackage{romannum}
\usepackage{xfrac}
\usepackage{float}
\usepackage{amsmath,bm}
\usepackage[title]{appendix}
\usepackage{todonotes}
\usepackage{subcaption}
\usepackage{caption}

\newtheorem{lemma}{Lemma}[section] 
\newtheorem{proposition}[lemma]{Proposition}
\newtheorem{design}[lemma]{Design}

\newtheorem{theorem}[lemma]{Theorem}

\newtheorem*{maintheorem*}{Main Theorem}
\theoremstyle{definition}{\newtheorem{definition}[lemma]{Definition}}

\newcommand{\rmf}{{\mathrm{f}}}

\newtheorem{rem}[lemma]{Remark}

\allowdisplaybreaks

\numberwithin{equation}{section}

\title[Control of plasma instabilities]{Control of Instability in a Vlasov-Poisson System Through an External Electric Field}
\date{\today}

\author[L. EINKEMMER]{Lukas Einkemmer}
\address[Lukas Einkemmer]{\newline Universitat Innsbruck, Innsbruck, Austria.}
\author[Q. Li]{Qin Li}
\address[Qin Li]{\newline University of Wisconsin Madison, Madison, USA.}
\author[C. Mouhot]{Clément Mouhot}
\address[Clément Mouhot]{\newline University of Cambridge, Cambridge, United Kingdom}
\author[Y. Yue]{Yukun Yue}
\address[Yukun Yue]{ \newline University of Wisconsin Madison, Madison, USA.}

\begin{document}
 \pagenumbering{arabic}
\maketitle

\begin{abstract}

Plasma instabilities are a major concern in plasma science, for applications ranging from particle accelerators to nuclear fusion reactors. In this work, we consider the possibility of controlling such instabilities by adding an external electric field to the Vlasov--Poisson equations. Our approach to determining the external electric field is derived from a linear analysis that examines the revised dispersion relation. Allowing the external electric field to depend on time and space, we show that it is possible to completely suppress the plasma instabilities when the equilibrium distribution and the perturbation are known, with one particular choice of external field turning the system back to free-streaming. Numerical simulations of the nonlinear two-stream and bump-on-tail instabilities verify our theory and demonstrate the effectiveness of the few choices of external electric field that we derive.

\end{abstract}

\section{Introduction}

Plasma, commonly recognized as the fourth state of matter, is an ionized gas composed of free electrons, ions, and neutral particles \cite{burm2012plasma,fitzpatrick2022plasma,frank2012plasma}. It is widely used in a variety of applications, and holds a promise of achieving green energy from fusion~\cite{ichimaru1993nuclear,miyamoto1980plasma}.

It is well known that plasma give rise to many instabilities, which often present significant challenges in both analysis and applications. Instabilities in plasma occur when perturbations, due to internal or external influences or imperfections, grow and disrupt an equilibrium state. In nuclear fusion, e.g., steady state operation in a homogeneous equilibrium is usually desired, but extremely challenging to achieve in a reactor (see, e.g, \cite{van2021beam, zohm2015magnetohydrodynamic}). As a consequence, to fully utilize plasma's capabilities, it is essential to study and manage these instabilities. Yet, this is immensely difficult and to today there is no systematic solution.

The fundamental model in studying microinstabilities (i.e.~instabilities due to kinetic as opposed to fluid behavior) is the Vlasov--Poisson (VP) equation \cite{chen2012introduction,nicholson1983introduction, schamel2000hole}, which we are going to use in this work. This equation is a fundamental tool that effectively models plasma as a collection of interacting charged particles, and traces the particle trajectories in a statistical manner. The equation is not only a physically useful model, due to the complicated features of its solutions, it has also become a focal point of extensive mathematical PDE research ~\cite{brenier2000convergence,guo2002vlasov,loeper2006uniqueness, mouhot2011landau, pfaffelmoser1992global,schaeffer1991global}. 

One intriguing property of the VP system is its interesting disparate behavior around equilibrium states. For example, a uniformly dense plasma with a Maxwellian velocity distribution is stable. A perturbation of such a system leads to the damping of electrostatic waves, known as Landau damping \cite{dawson1961landau,mouhot2011landau,ryutov1999landau}. However, if the velocity distribution is modified one generally observes instabilities. Two important examples of such instabilities that are modeled by the Vlasov--Poisson equation are the two-stream instability and the bump-on-tail instability. The onset of these instabilities is well described by linear theory and consequently such results are commonly used in the physics literature to predict whether a given instability is present in a system.

The two-stream instability emerges when two streams of charged particles within a plasma move at very different velocities (sometimes opposite to each other). This phenomenon is prevalent in contexts such as electron beams traversing through plasma \cite{lapuerta2002general,mcbride1972theory, roberts1967nonlinear} or in particle accelerators \cite{chen1986energy}. An initial perturbation present in the beams is amplified and leads to usually unwanted growth of the perturbation and energy transfer from the beams to the waves. This can culminate in beam scattering and focus loss, posing substantial challenges in both plasma experiments and applications \cite{sydorenko2016effect}.

The Bump-on-tail instability is another instability that has attracted a lot of study. This phenomenon is commonly observed when an energetic particle beam is introduced into a thermalized plasma. The instability disrupts the equilibrium and fosters Langmuir wave growth \cite{dum1990simulation,goldman1996langmuir}. This instability is particularly significant for nuclear fusion research as it is one of the simplest models for the bump-on-tail features generated in fusion reactors by either neutral beam heating \cite{speth1989neutral} or runaway electrons \cite{breizman2019physics}.

The timescale on which microinstabilities take place can be extremely rapid: often in the range of nanoseconds. At this timescale, there is very limited information on the velocity distribution available. So feedback-based control, even though mathematically well understood \cite{doyle2013feedback,skogestad2005multivariable}, can be extremely challenging to implement for plasma systems (some effort has been put in this direction~\cite{albi2024instantaneous}). However, laboratory experiments and computer simulations seem to indicate that time-dependent perturbations in the plasma can have a stabilizing effect (see, e.g., \cite{kawata2019dynamic}). Recently, in \cite{Einkemmer2023}, it has been shown (by numerical optimization) that even a time-independent external electric field can delay the onset of a two-stream instability for the Vlasov--Poisson equations.

While these results indicate the possibility of suppressing plasma instabilities by applying external electric fields, to the best of the authors' knowledge, there is currently no analysis available. In this work, we provide a linear stability analysis for the Vlasov--Poisson equation with an external electric field. Specifically, assuming the initial state of the plasma is known, we are able to:

\begin{itemize}
   \item Formulate explicit expressions for a class of external fields that can suppress instabilities emerging from arbitrary unstable equilibrium states;
   \item We apply this theory to find electric fields that completely suppress the two-stream and bump-on-tail instabilities (without the use of feedback control). The numerical simulation confirms the theoretical prediction.
   \item One approach devised from the framework recovers a free-streaming solution. In this situation, the plasma acts as if there is no electrostatic force and damping is assured.
\end{itemize}

{It is worth noting that, in contrast to the results reported in \cite{Einkemmer2023}, where the effectiveness of suppressing instability relies heavily on selecting an appropriate initial condition, the strategy proposed in this work incorporates time information to provide an explicit formula directly, eliminating the need for an optimization process to determine a suitable external field. Furthermore, while the approach in \cite{Einkemmer2023} can only suppress instability up to a certain time, some of the strategies presented in this work are capable of suppressing instability for all \( t > 0 \).
}

The core of our study relies on the linear analysis strategy established as early as Landau \cite{Landau1965}. These early works link the local-in-time dynamics around equilibrium states with the pole analysis of the PDE operator. Depending on the signs of the poles, the perturbation either exponentially grows or damped. For the system that demonstrates exponential growth, our strategy essentially is to find external electric fields that can cancel out these poles, so as to counter-balance the growth. 

{We would like to clarify that the aim of this work is to provide a theoretical foundation for exploring potential control strategies to mitigate instabilities in plasma, rather than proposing a directly implementable experimental design. The external electric field considered in this study incorporates both spatial and temporal components. In practical applications, high-intensity lasers can be used to influence plasma dynamics. Such laser drives can induce nonlinear responses in the plasma, potentially leading to the formation of kinetic electrostatic electron nonlinear (KEEN) waves, as discussed in \cite{Afeyan2014, cheng2013study} and references therein. This form of control is close to our formulation. On the other hand, implementing control through a magnetic field is the most experimentally feasible approach. To achieve this, one could either perform numerical optimization \cite{albi2024instantaneous} or conduct a root analysis similar to the one presented in this work. Ultimately, this study serves as a foundational step toward future investigations under more practical settings.
}

The structure of this paper is outlined as follows: In Section \ref{sec:vlasov-poisson}, we provide a concise review of the VP system from a mathematical perspective. This section includes a detailed introduction to the evolution of small perturbations in different equilibrium states and presents the standard linear analysis of the VP system using Fourier and Laplace transforms. Additionally, we introduce the Penrose condition in this section. Following this, Section \ref{sec:control} and Section \ref{sec:negating_electric_field} detail our main contributions, presenting two strategies for controlling plasma instabilities. The method described in Section \ref{sec:control} extends the linear analysis to the VP equation with an external field, providing a general framework to guide the design of an external field to suppress instability. The second method, described in Section \ref{sec:negating_electric_field}, is based on a physical idea to negate the self-generated electric field with an external field. We demonstrate that this second method is essentially a special case within the broader framework of the first. Section \ref{sec:numerical} showcases numerical evidence supporting the effectiveness of our proposed methods when applied to two-stream and bump-on-tail instabilities.

\section{Vlasov-Poisson equation}\label{sec:vlasov-poisson}
In this section we first review some basic properties of the VP system and discuss the stability analysis around equilibrium states.
\subsection{Mathematical Model}
The VP system is a fundamental model used in simulating dynamics of plasma. When incorporates an external electric field \( H \), the system is described as~\cite{DiPerna1988,Dolbeault1991}:
\begin{equation}\label{eq:VP_system}
\left\{
    \begin{aligned}
        &\partial_t f(t,x,v) + v \cdot \nabla_x f(t,x,v) - \left[E(t,x) + H(t,x)\right] \cdot \nabla_v f(t,x,v) = 0, \\
       &\Delta V(t,x) = 1 - \rho(t,x), \\
       &E(t,x) = \nabla V(t,x),
    \end{aligned}
\right.
\end{equation}
where \( f(t,x,v) \) is the phase-space distribution function, describing the particle density at position \( x \), velocity \( v \), and time \( t \). Here, the spatial variable lies in the periodic box $\mathbb{T}^d$ and the velocity variable $v \in \mathbb{R}^d$. We consider the torus here as our spatial domain because Tokamaks \cite{ariola2008magnetic,wesson2011tokamaks}, and other important devices are constructed to impose periodic boundary conditions. In time, particles moves with velocity $v$ and are accelerated according to the total electric field $E(x,t)+H(t,x)$. Here $H$ is the external electric field, and \( E \) is the self-generated field given by the gradient of the self-generated electric potential $V$, which is determined by the charged density $\rho(t,x)$ through the Poisson equation:
\begin{equation}\label{eq:rho}
    \rho(t,x) = \int f(t,x,v) \, dv. 
\end{equation}

When the initial data is prepared as a spatially-independent function, $\nabla_xf=0$ and $E=\nabla V=0$, and the system does not evolve. This means all functions that does not have spatial dependence, denoted $\mu(v)$, are equilibrium states.

While some equilibrium states are stable, others are not. We are primarily concerned with equilibria that are not stable and magnify the electric field when a small perturbation is introduced. {Decomposing} \( f(t,x,v) = f_{\text{p}}(t,x,v) + \mu(v) \), with \( f_{\text{p}} \) representing the component of the dynamics that responds to the perturbation, then:
\begin{equation}\label{eq:VP_system_pert}
\left\{
    \begin{aligned}
        &\partial_t f_{\text{p}}(t,x,v) + v \cdot \nabla_x f_{\text{p}}(t,x,v) - \left[E(t,x) + H(t,x)\right] \cdot \nabla_v \left[f_{\text{p}}(t,x,v) + \mu(v)\right] = 0, \\
       &\Delta V(t,x) = -\rho_{\text{p}}(t,x), \\
       &E(t,x) = \nabla V(t,x).
    \end{aligned}
\right.
\end{equation}

In the derivation, the fact \( \partial_t \mu + v \cdot \nabla_x \mu = 0 \) has been utilized, and we used the notation \( \rho_{\text{p}}(t,x) \) to represent the perturbed charge density:
\begin{equation}
    \label{eq:rho_p}
      \rho_{\text{p}}(t,x) = \int f_{\text{p}}(t,x,v) \, dv.
\end{equation}

{Moreover, in the linearized regime,} with $f_\text{p}\ll\mu$ in a proper metric, the nonlinear term is dropped, resulting in the linearized equation:
\begin{equation}\label{eq:VP_system_pert_linear}
\left\{
    \begin{aligned}
        &\partial_t f_{\text{p}}(t,x,v) + v \cdot \nabla_x f_{\text{p}}(t,x,v) - \left[E(t,x) + H(t,x)\right] \cdot \nabla_v \mu(v) = 0, \\
       &\Delta V(t,x) = -\rho_{\text{p}}(t,x), \\
       &E(t,x) = \nabla V(t,x)\,.
    \end{aligned}
\right.
\end{equation}

Throughout this paper, our analysis primarily centers on the systems defined in \eqref{eq:VP_system_pert} and its linearized approximation \eqref{eq:VP_system_pert_linear}. For the sake of clarity and simplicity in our exposition, we drop, in the following, the subindex $\text{p}$ and present derivation for the 1D problem. Consequently, the full distribution will be presented as \( f + \mu \). {It is worth noting that if $E(t,x) + H(t,x) = 0$ for all values of $t$ and $x$ in equation \eqref{eq:VP_system_pert} or \eqref{eq:VP_system_pert_linear}, the Vlasov-Poisson equation reduces to the free-streaming equation. This observation relates to one of the suppression strategies we propose, which will be discussed in further detail in Section \ref{sec:negating_electric_field}.
}

\subsection{Stability of equilibrium}
As has been mentioned, a state of equilibrium can be disrupted by the introduction of a small perturbation. This perturbation triggers the self-generated electric field that then accelerates the particles in the plasma.

Sometimes the self-generated field cancels out the effect of the perturbation, and the state automatically damps back to the equilibrium state, but in other times the self-generated field further magnifies the perturbation. The difference leads to the distinction between stable and unstable equilibrium. Here we provide three most well-known examples, which have been mentioned in the introduction, to illustrate such phenomenon:
\begin{itemize}
    \item The Maxwellian distribution is a standard Gaussian distribution and represents a stable equilibrium. A small perturbation to the field will be naturally damped out by the system, a phenomenon known as Landau damping.
    \item The two-stream distribution consists of a combination of two Gaussian distributions with equal weights, illustrating a scenario where two distinct streams of particles coexist. This is an unstable equilibrium, where a small perturbation to the field can be exponentially amplified \cite{nicholson1983introduction}.

    \item The bump-on-tail distribution is a composite of two Gaussian distributions with different weights, where one distribution is notably larger than the other. This describes the situation where a small group of particles moving at very high speeds is suddenly introduced into an equilibrium consisting primarily of particles sampled from a standard Gaussian distribution. In this case, the high-speed particles create a small ``bump" on the Gaussian tail \cite{berk1995numerical}. This is also an unstable equilibrium.
\end{itemize}

\begin{figure}
    \centering
    \includegraphics[width=6.8in]{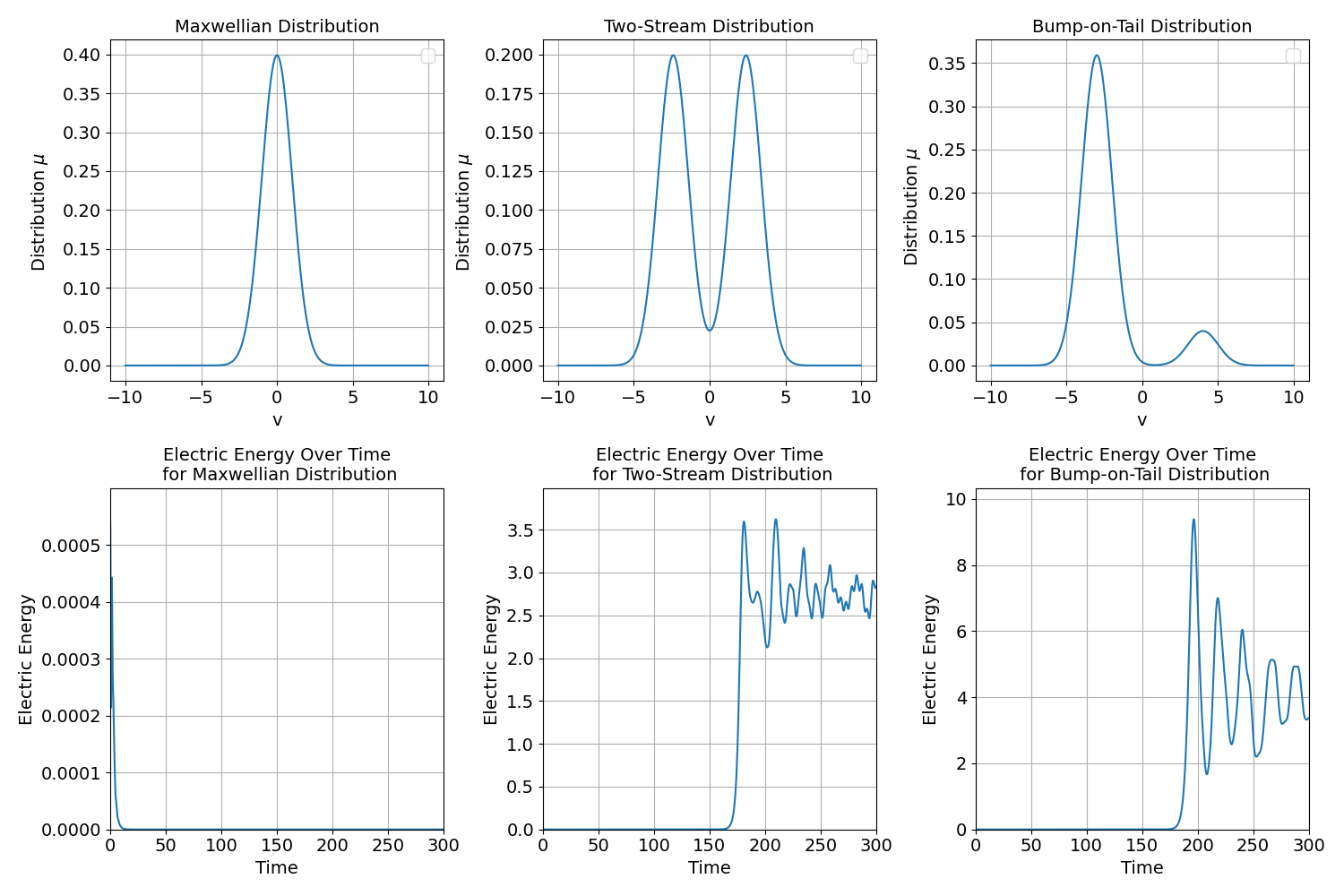}
    \caption{Dynamics of the electric energy in response to a uniform small perturbation in the VP system without an external field.
}
\label{fig:electric_energy_different_distribution}
\end{figure}

Introducing a small-amplitude perturbation to these three distinct equilibrium states might lead to different outcomes. In general, a Maxwellian distribution will be stable against perturbations, while a two-stream distribution and a bump-on-tail distribution with large enough bumps might be unstable \cite{mouhot2011landau,penrose1960electrostatic}. This variation is illustrated in Figure \ref{fig:electric_energy_different_distribution}, where those distributions, despite receiving the identical perturbation, exhibits various responses. Specifically, in the Maxwellian distribution scenario, the wave exhibits rapid damping. Conversely, for the specific two-stream and bump-on-tail distributions presented in the second and the third column of Figure \ref{fig:electric_energy_different_distribution}, the same perturbation induces a marked increase in wave activity, {leading to significant growth of electric energy. This observation confirms that} a Maxwellian distribution typically results in Landau damping of electrostatic waves \cite{dawson1961landau,mouhot2011landau,ryutov1999landau}, while starting from a two-stream or bump-on-tail equilibrium with large enough bumps structures leads to exponential amplifications of wave energies. While we refrain from delving into the numerical specifics here, readers are directed to Section \ref{sec:numerical} for further details. It is noteworthy to highlight that this phenomenon is not novel and has been investigated since the fundamental work by Landau \cite{Landau1965}. The Penrose condition \cite{penrose1960electrostatic} serves as a physical criterion for predetermining equilibrium stability prior to any numerical simulation.

Our goal here is to stabilize the system by applying an external electric field \(H\). As previously mentioned, we plan to introduce two approaches to mitigate these instabilities. The specifics of these methods will be discussed in the subsequent Sections~\ref{sec:control} and \ref{sec:negating_electric_field}. In~\ref{sec:LinearPenrose} we provide a quick review of the classical linear analysis for the VP system without the external field. This is the foundation of our further studies.

\subsection{Linear Analysis}\label{sec:LinearPenrose}

The different stability behaviors mentioned above can be explained through linearized studies of the VP system, with the linearization conducted around the equilibrium states. This is a commonly used strategy that has been deployed in various studies \cite{Adkins2017, bedrossian2020nonlinear, bedrossian2016landau, bedrossian2018landau, HanKwan2021}, followed from the fundamental work \cite{mouhot2011landau}. Many results revolve around resolvent (or root) estimates.

We use the symbol \(\hat{X}\) to denote the Fourier transform of a function \(X\) {with respect to the spatial $x$ (and possibly) velocity variable $v$}, and $L$ stands for the Laplace transform:
\begin{equation}
    \label{eq:Laplace_transform_def}
    L[F](s)=\int_0^\infty e^{-st }F(t)\,dt\,.
\end{equation}

As expected, in the linear regime, the equation's behavior is solely determined by its spectrum. Indeed, when the external field $H=0$, we can compute the relation between the Laplace and Fourier transform of certain quantities:
\begin{equation}   \label{eq:external_Laplace_no_external_field}
     L[\hat{\rho}(\cdot,k)](s)\left( 1+L[\hat{U}(\cdot,k)](s) \right) = L[\hat{S}(\cdot,k)](s)\,.
\end{equation}

The two involved notations are the following:
\begin{itemize}
\item \( \hat{U}(t,k) \) is defined as:
\begin{equation}\label{eq:hatU_def}
 \hat{U}(t,k) :=   t\hat{\mu}(kt)\,.
\end{equation}
It is a function of \( t \) and \( k \), and is uniquely determined by the equilibrium \(\mu\).
\item \( \hat{S} \) is defined as
\begin{equation}\label{eq:hatS_def}
    \hat{S}(t,k) := \hat{f}(0,k, kt) \,,
\end{equation} 
where $\hat{f}$ takes the Fourier transform in both $x$ and $v$. According to this definition, \(S\) is determined solely by the initial value of the perturbation. This definition has a very nice physical interpretation. Suppose $f_\rmf$ only propagates as a free-stream (neglecting electric fields),
\begin{equation}\label{eq:drift_equation}
    \partial_t f_\rmf + v\partial_x f_\rmf = 0\,,
\end{equation}
with the subindex denoting it is a free-stream solution, then the solution is
\( f_\rmf(t, x, v) = f_0(t, x - vt, v) \). Denoting $\rho_\rmf = \int f_\rmf dv$ as usual, we notice
\begin{equation}\label{eqn:s_interpretation}
\hat{S}(t,k)=\hat{\rho}_\rmf(t,k)=\hat{f}(0,k, kt)\,.
\end{equation}
\end{itemize}

The significance of the equation~\eqref{eq:external_Laplace_no_external_field} is immediate. {It provides a very simple form that relates the to-be-solved quantity $\rho$ with the initial data coded in $S$ and the equilibrium state coded in $U$.} The solvability condition for $\hat\rho$ now purely depends on the roots of $1+L[\hat{U}]$:
\begin{itemize}
    \item If for all $k$, the factor $1+L[\hat{U}]$ is away from $0$ for all $\mathcal{R}s>0$, then $\rho$ can be uniquely solved by
    \[
    L[\hat{\rho}(\cdot,k)](s) = \frac{L[\hat{S}(\cdot,k)](s)}{ 1+L[\hat{U}(\cdot,k)](s)}\,.
    \]
    \item If for some $k$, the factor $1+L[\hat{U}]$ has a root $s$ with positive real component, there can be an exponentially growing mode of $\rho$ coded by this root.
\end{itemize}
This insight serves as the foundation for the classical Penrose stability condition \cite{HanKwan2021,penrose1960electrostatic}:
\begin{definition}
    An equilibrium distribution function \(\mu(v)\) is said to satisfy the Penrose stability condition if there exists a $\kappa_0 >0$ such that
    \begin{equation*}
        \inf_{k \in \mathbb{Z},\, s \in \mathbb{C},\, \mathcal{R}s\geq 0} \left| 1 + L[\hat{U}(\cdot,k)](s) \right| \geq \kappa_0.
    \end{equation*}
    where \(\hat{U}(t,k)\) is defined in \eqref{eq:hatU_def} and $\mathcal{R}s$ denotes the real part of $s$.
\end{definition}

It is important to note that the Gaussian distribution meets the Penrose condition, leading to the famous Landau damping \cite{mouhot2011landau,nicholson1983introduction}. However, for the two-stream distribution or bump-on-tail distribution, the Penrose condition is not met, meaning the term \( \left( 1+L[\hat{U}(\cdot,k)](s) \right) \) has positive roots, which suggests instability might occur.

\begin{figure}
    \centering
    \includegraphics[width=7in]{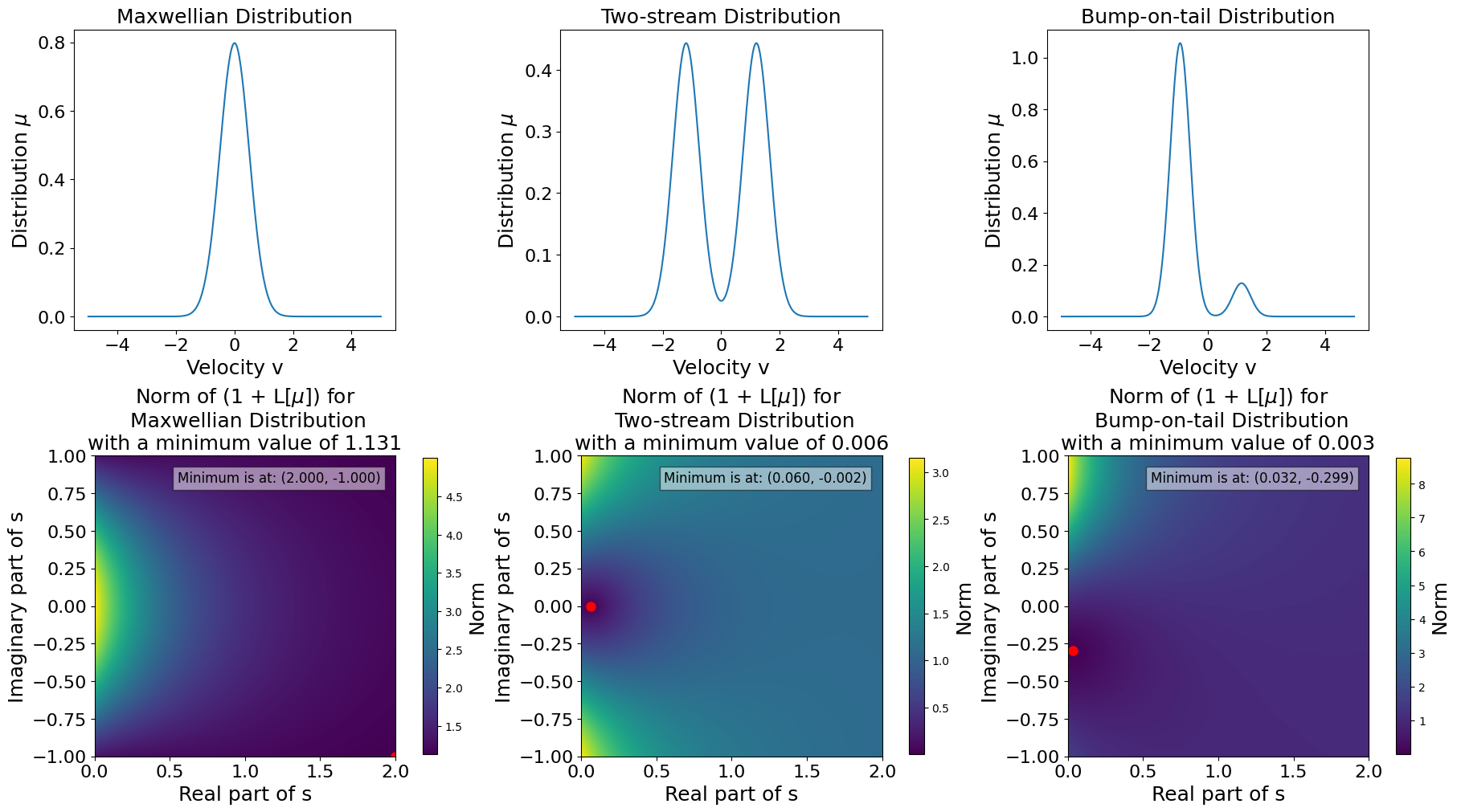}
    \caption{Norm of $\left(1+L[\hat{U}(\cdot,1)](s)\right)$ for different equilibrium functions for $k=1$. {Red dots are the roots found numerically. For two-stream and bump-on-tail distributions, roots are found on the half plane where $\mathcal{R}s>0$, signaling the potential exponential growth of electric field.}}
    \label{fig:1+L[f]}
\end{figure}

Figure \ref{fig:1+L[f]} shows this difference clearly. Here, we plot the magnitude of \( \left( 1+L[\hat{U}(\cdot,k)](s) \right) \), looking at three different distributions as the equilibrium $\mu$. We have chosen to plot the factor for \(k=1\).

\begin{itemize}
    \item \textbf{Gaussian:} The minimum value of \(\left(1 + L[\hat{U}(\cdot, 1)](s)\right)\) always stays above 1 and roots have not been found in the positive part of $s$. This observation is in line with the well-established understanding that the Gaussian distribution fulfills the Penrose condition.
    
    \item \textbf{Two-stream:} Dark blue regions in the two-stream plot show areas where \(\left(1 + L[\hat{U}(\cdot, 1)](s)\right)\) decreases towards zero. A root $s$ with positive real component has been found{, as presented by the red dots in Figure~\ref{fig:1+L[f]}}. This validates the fact that two-stream initial condition does not meet the Penrose condition and is unstable.
    
    \item \textbf{Bump-on-tail:} Similarly, we found roots with $\mathcal{R}s>0$ for the bump-on-tail scenario, meaning the initial condition fails to satisfy the Penrose condition, and is unstable.
\end{itemize}

\section{Pole elimination: An Approach Through A Linear Analysis}
\label{sec:control}
We will now extend the classic linear stability analysis in section \ref{sec:LinearPenrose} to the Vlasov--Poisson equation with an external electric field $H$.
\begin{proposition}
    \label{prop:external_Laplace}
    In the presence of an external field \(H\), assume \(f\) solves the Vlasov-Poisson system \eqref{eq:VP_system_pert_linear} and \(\rho\) is defined according to \eqref{eq:rho_p}. The following identity holds:
    \begin{equation}
     \label{eq:external_Laplace}
     \left(L[\hat{\rho}(\cdot,k)](s)-L[\hat{S}(\cdot,k)](s)\right)\left( 1+L[\hat{U}(\cdot,k)](s) \right) = L[\hat{U}(\cdot,k)](s)\left(-L[\hat{S}(\cdot,k)](s) + ik L[\hat{H}(\cdot,k)](s)  \right).
\end{equation}
\end{proposition}

The derivation is straightforward, and we leave it to appendix \ref{appen:proof_external_Laplace}. 

\begin{rem}
    \label{rem:Laplace_Fourier_form_original}
    The formula~\eqref{eq:external_Laplace} may be seemingly complex at first. In fact, the direct application of Fourier and Laplace transforms results in a formula where the term \( L[\hat{S}(\cdot,k)]L[\hat{U}(\cdot,k)] \) cancels out on both sides. This simplification, as demonstrated in the proof of Proposition \ref{prop:external_Laplace} in Appendix \ref{appen:proof_external_Laplace}, reduces equation \eqref{eq:external_Laplace} to:
\[
 L[\hat{\rho}(\cdot,k)](s)\left( 1 + L[\hat{U}(\cdot,k)](s) \right) = L[\hat{S}(\cdot,k)] + ik L[\hat{U}(\cdot,k)](s)L[\hat{H}(\cdot,k)](s) \,.
\]
{We nevertheless opt to operate on the seemingly more complex equation~\eqref{eq:external_Laplace} because it delivers more physics intuition. More specifically, this equation allows us to examine the deviation of the to-be-solved solution $\rho$ and the desired free-streaming solution $S$.}
\end{rem}

Comparing with~\eqref{eq:external_Laplace_no_external_field}, it is clear that the newly added $H$ merely adds another source term on the right hand side. From this point of view, stability can be achieved by choosing $H$ {that cancels the root effect from the factor on the left hand side.} As discussed in Section~\ref{sec:LinearPenrose}, the stability of $\rho$ mainly depends on the root condition of $ 1 + L[\hat{U}(\cdot, k)] $ when no external field presents. In scenarios where there are positive roots that lead to exponential growth of electric field, one natural strategy is to design $H$ so that the factor term on the right of~\eqref{eq:external_Laplace} cancels out the roots of $ 1 + L[\hat{U}(\cdot, k)] $. By doing so, we obtain an explicit formula for $L[\hat{\rho}(\cdot,k)](s)-L[\hat{S}(\cdot,k)](s)$. This strategy will be henceforth referred to as ``Pole Elimination''.

\begin{design}[Pole Elimination]
    \label{dsn:external_Laplace}
    We are to choose \( \hat{H} \) to satisfy:
\begin{equation}
    \label{eq:external_field_choice}
    -L[\hat{S}(\cdot,k)](s) + ik L[\hat{H}(\cdot,k)](s) = h\left( 1 + L[\hat{U}(\cdot,k)](s) \right),
\end{equation}
where the function $h$ should satisfy $h(0) = 0$ and
\begin{equation}\label{eqn:h_cond}
    h : \mathbb{C} \to \mathbb{C}\quad\text{s.t.}\quad \lim\limits_{|x| \to 0} \frac{|h(x)|}{|x|} = c \ \ \text{for some} \ c\in\mathbb{C}\,.
\end{equation}
For this $H$, {equation~\eqref{eq:external_Laplace} has a closed form solution:}
\[
L[\hat{\rho}(\cdot,k)](s)=L[\hat{S}(\cdot,k)](s)+L[\hat{U}(\cdot,k)](s)\frac{h\left( 1 + L[\hat{U}(\cdot,k)](s) \right)}{1 + L[\hat{U}(\cdot,k)](s)}\,,
\]
and close to the roots, the solution is a linear combination of free-streaming and equilibrium state $\hat{\rho}=\hat{S}+c\hat{U}$.
\end{design}

In the rest of this section, we study multiple specific examples of $h(x)$, all of which satisfy the condition~\eqref{eqn:h_cond}, and we investigate the corresponding $H$ and the associated behavior of the solution.

\subsection{Selecting ${h(x)=0}$:}\label{sec:h_0}

The easiest choice is to make the entire right-hand side of \eqref{eq:external_Laplace} identically zero by choosing \( h(x) = 0 \), then:
\begin{equation*}
    -L[\hat{S}(\cdot,k)](s) + ik L[\hat{H}(\cdot,k)](s) = 0\,,
\end{equation*}
and from~\eqref{eq:external_Laplace} we then follow:
\[
L[\hat{\rho}(\cdot,k)]=L[\hat{S}(\cdot,k)]\,.
\]
Applying the inverse Laplace transform:
\begin{equation*}
    ik\hat{H}(t, k) = \hat{S}(t, k) = \hat{\rho}(t,k)\,,
\end{equation*}
and follow the argument in~\eqref{eq:drift_equation} and~\eqref{eqn:s_interpretation}, we have:
\begin{equation}
    \label{eq:h(x)=0_H}
    ik\hat{H}(t, k) = \hat{S}(t, k) = \hat{\rho}(t,k) = \hat{f}(0, k, kt) = \hat{\rho}_\rmf(t, k)\,.
\end{equation}
Two equalities are interesting:
\begin{itemize}
\item From $ik\hat{H}(t, k) = \hat{\rho}_\rmf(t, k)$, we take the inverse Fourier transform to have:
\begin{equation}
    \label{eq:H_choice_h(x)=0}
    H(t,x)=\nabla \Delta^{-1}\rho_\rmf(t,x) = -E_\rmf(t,x)\,,
\end{equation}
where $E_\rmf$ is the electric field generated by the free-stream density $\rho_\rmf$. This means the external field $H$ is set to \underline{counter-balance} the free-streaming field.
\item From $\hat{\rho}(t,k) = \hat{\rho}_\rmf(t, k)$, we obtain that the VP system with external electric field $H$ defined in~\eqref{eq:H_choice_h(x)=0} \underline{exactly recovers} the free-streaming solution.
\end{itemize}
All together, these observations suggest that this choice of $h=0$ brings a rather interesting phenomenon: if the external electric field completely counteracts the free-stream field, the VP system is reverted to the free-streaming system, and is stabilized. Due to the effect of this approach, we also term this approach ``Electric Field Neutralization" and will discuss it in more detail in Section \ref{sec:negating_electric_field}.

\subsection{Selecting \({h(x)=\alpha x}\)} \label{sec:hx=alphax}
Another choice of $h(x)$ that allows analytical computation is to set it as a linear function $h(x)=\alpha x$ with $\alpha$ being a weight parameter.

In this case, using
\eqref{eq:external_field_choice}, we have:
\begin{equation*}
    -L[\hat{S}(\cdot,k)](s) + ik L[\hat{H}(\cdot,k)](s) = \alpha \left( 1 + L[\hat{U}(\cdot,k)](s) \right).
\end{equation*}
With the inverse Laplace transform:
\begin{equation}\label{eq:h(x)=x_choice}
    ik\hat{H}(t, k) = \hat{S}(t, k) + \alpha\left(\delta(t) + \hat{U}(t, k) \right),
\end{equation}
where \( L^{-1}\{1\} = \delta(t) \)~\cite{fox2002mathematics}. The explicit formula for $H$ can be obtained by taking the inverse Fourier transform. To see its effect on $\rho$, we insert it back into~\eqref{eq:external_Laplace}, then
\begin{equation*}
     \left(L[\hat{\rho}(\cdot,k)](s)-L[\hat{S}(\cdot,k)](s)\right)\left( 1+L[\hat{U}(\cdot,k)](s) \right)=\alpha L[\hat{U}(\cdot,k)](s)\left( 1+L[\hat{U}(\cdot,k)](s) \right).
\end{equation*}
Assuming that we can cancel out
\( \left( 1 + L[\hat{U}(\cdot,k)](s) \right) \) on both sides (which we do not prove rigorously here), we have
\begin{equation}\label{eq:rho_h(x)=x}
    \hat{\rho}(t, k) = \hat{S}(t, k) + \alpha \hat{U}(t, k)\,.
\end{equation}
Or equivalently, denoting $S$ and $U$ the inverse Fourier transform,
\begin{equation}\label{eq:rho_h(x)=x_physical}
    {\rho}(t, x) = {S}(t, x) + \alpha {U}(t, x)\,.
\end{equation}
We should note that $S(t,x)=\rho_\rmf$ still has a physical meaning, see discussion in~\eqref{eqn:s_interpretation}, while $U(t,x)$ is a quantity appeared through mathematical manipulation that does not necessarily carry a physical meaning. The two terms respectively code the initial perturbation and the equilibrium state. Since both terms can be pre-computed, we can use this knowledge to conduct strategic selection of \(\alpha\). When $\alpha=0$, we recover the situation in Section~\ref{sec:h_0}. Using the information about $U$ and $S$ to tune $\alpha$ provides another layer of flexibility for our control. We present one example each for two-stream instability and bump-on-tail instability in Section \ref{sec:numerical}.

\subsection{Remarks on other possibilities}\label{subsec:remarks_h_choices}

There are other possibilities that satisfy the requirement of~\eqref{eqn:h_cond}. One obvious choice is to set $h(x)=x^2$. This would lead to a scenario of
\[
-L[\hat{S}(\cdot,k)](s) + ik L[\hat{H}(\cdot,k)](s) = \left( 1 + L[\hat{U}(\cdot,k)](s) \right)^2\,.
\]
This also removes the poles and brings stability to the problem. However, to obtain $H$ we have to numerically compute an inverse Laplace transform (in contrast to the two previous choices where this can be done analytically). The high numerical sensitivity of the Laplace and inverse Laplace transform can cause loss of accuracy, see~\cite{cohen2007numerical,craig1994practical}. For this reason, {this particular choice of $h$ will not be investigated numerically.}

Furthermore, the choice of \( h \) does not have to be preset. Viewing \( h(x) = 0 \) as a special case of \( h(x) = \alpha x \), we can set \( \alpha(t) \) as a function of time and adjust it dynamically. Indeed, this is what we will use in Section \ref{sec:numerical} as the most effective method, where we implement a two-phase control. Initially, we use \( h(x) = \alpha x \) with a constant \( \alpha \) to rapidly damp the instabilities, and then transition to \( h(x) = 0 \) to maintain equilibrium. This approach allows us to benefit from reduced oscillations and a quick decay rate simultaneously.

\section{Electric Field Neutralization}\label{sec:negating_electric_field}

We study in depth the Electric Field Neutralization, the strategy proposed in Section~\ref{sec:h_0}. The earlier derivation suggests that by designing $H$ to completely counter-balance the electric field generated by free-streaming, the VP system is reverted to the simple free-streaming solution. We present more details in this section. In particular, we will show that this statement not only holds true in the linearized regime, as discussed in Section \ref{sec:control}, but is true in nonlinear setting as well. We will also demonstrate that free-streaming is indeed stable.

We recall the free-streaming system, as already presented in~\eqref{eq:drift_equation}:
\begin{equation}\label{eq:drift_equation_mod}
    \partial_t f_\rmf + v\cdot\nabla_x f_\rmf = 0.
\end{equation}
Assuming that the initial data  \( f(0,x, v) \) is known, the solution to this free-streaming system can be obtained as \( f_\rmf(t, x, v) = f(0, x - vt, v) \). Consequently,
\begin{equation}
    \label{eq:E_neutralization}
    E[f_\rmf](t,x) = \nabla \Delta^{-1}\left( \int_{\mathbb{R}} f(0, x - vt, v)\,dv \right)\,.
\end{equation}

We claim that by setting $H$ to be negative of the electric field generated by $\rho_\rmf$, the original VP system~\eqref{eq:VP_system_pert} recovers the free-streaming~\eqref{eq:drift_equation_mod}.
\begin{lemma}\label{lem:neutralization}
    Let $f_\rmf(t,x,v)$ solve~\eqref{eq:drift_equation_mod} and $f_2(t,x,v)$ be the solution to the VP system~\eqref{eq:VP_system_pert} with $H=-E[f_\rmf]$. If they have the same initial data, then $f_\rmf=f_2$ for all time.
\end{lemma}
\begin{proof}
We first write the equation for $f_2$: 
\begin{equation}\label{eq:VP-neutral}
\left\{
    \begin{aligned}
        &\partial_t f_2 + v \cdot \nabla_x f_2 - (E[f_2] - E[f_\rmf])\cdot \nabla_v (f_2+\mu) = 0, \\
        &\Delta \left(V[f_2]\right) = -\int_{\mathbb{R}} f_2\,dv,\\
        &E[f_2] = \nabla \left(V[f_2]\right). 
    \end{aligned}
\right.
\end{equation}
Define $f^\delta=f_2-f_\rmf$ then we know $f^\delta$ satisfies: 
    \begin{equation*}
        \partial_t f^\delta+v\cdot\nabla_x f^\delta - (E[f_2]-E[f_\rmf])\cdot\nabla_v (f_2+\mu)=0.
    \end{equation*}
    By linearity of $E[f]$, this equation can be rewritten as
    \begin{equation}
        \label{eq:fdelta_equation}
          \partial_t f^\delta+v\cdot\nabla_x f^\delta - E[f^\delta]\cdot\nabla_v (f_2+\mu)=0
    \end{equation}
Since $f^\delta(0,x,v)=0$, $f^\delta(t,x,v)=0$ is the unique trivial solution to \eqref{eq:fdelta_equation}, implying that $f_2=f_\rmf$ for all time. For more details on the derivation of uniqueness, we refer readers to \cite[Theorem 5.2.1]{glassey1996cauchy}.

\end{proof}

Thus, we have shown that for this particular choice the solution of the nonlinearized Vlasov--Poisson equation with external electric field degenerates to the simple free-streaming problem. The choice we have made for the external electric field, i.e. $H_1 = -E[f_1]$, is precisely what comes out of our analysis in the previous section, i.e.~is identical to equation \eqref{eq:H_choice_h(x)=0}.

Recovering free-streaming immediately leads to stability. Indeed, for many practical initial data (whose Fourier transforms decay super-exponentially in Fourier space \cite{mouhot2011landau}, including two-stream and bump-on-tail instabilities), the density and electric field to the free-streaming problem decays to zero superexponentially in time:

\begin{proposition}

    \label{prop:drift_exponential_decay}
  Let $f_\mathrm{f}(t,x,v)$ solve \eqref{eq:drift_equation_mod}, 
    the free-streaming equation. Assume \( f(0,x,v) = \mu(v)X(x) \), where \( \mu \) is a Gaussian mixture (a linear combination of Gaussian distributions) and \( X \) is a periodic function,with $\hat{X}(0) = 0$, where $\hat{X}$ denotes the Fourier transform of $X$. Then the charge density function \( \rho_\rmf(t,x) = \int_{-\infty}^{\infty} f_\rmf(t,x,v) \, dv \) exhibits the following decay:
    $$|\rho_\rmf(t,x)|<Ce^{-At^2},$$
    pointwisely in $x$, where $C$ and $A$ are constants that only depend on the initial perturbation.
\end{proposition}

The detailed proof of this result is presented in Appendix \ref{appen:proof_drift_decay}. Built upon these findings, we arrive at the central theorem of this section:

\begin{theorem}
    \label{thm:neutralization_result}
   Assume that $H = -E[f_\rmf]$, the initial condition is of the form described in Proposition \ref{prop:drift_exponential_decay}, with the equilibrium distribution $\mu(v)$ being a Gaussian mixture, then the solution to Vlasov-Poisson system \eqref{eq:VP_system_pert} demonstrates stability in its electric energy. More precisely, the electric energy $\mathcal{E}(t)$ exhibits exponential decay, satisfying:
   \begin{equation*}
       \mathcal{E}(t) < \Tilde{C}e^{-At^2},
   \end{equation*}
where $\mathcal{E}(t)$ is defined as $\mathcal{E}(t) = \tfrac{1}{2}\int  E(t,x)^2\,dx$.
Here, $\Tilde{C}$ is a constant depending on the initial conditions and the domain, and $A$ is the same constant identified in Proposition \ref{prop:drift_exponential_decay}.
\end{theorem}

\begin{proof}
  Noting that the average value of $V$ does not affect the solution of the system \eqref{eq:VP_system_pert}, we can assume without loss of generality that $V$ satisfies $\int_X V(t,x)\,dx = 0$. Utilizing the standard regularity theory for elliptic equations \cite{brezis2010functional,gilbarg1977elliptic}, the solution to the Poisson equation in \eqref{eq:VP_system_pert} satisfies
    \begin{equation*}
        \mathcal{E}(t)= \frac{1}{2}\lvert V(t,\cdot)\rvert_{H^1(X)}^2 \leq C \|\rho(t,\cdot)\|_{L^2(X)}^2 \leq C {m(X)}\,\left[\max_{x \in X} |\rho(t,x)|\right]^2,
     \end{equation*}
    where $H^1(X)$ represents the Sobolev space defined on domain $X$, $m(X)$ is the Lebesgue measure of domain $X$ which is assumed to be finite, and $\lvert\cdot\rvert_{H^1(X)}$ denotes the semi-norm in this space. Then the result we aim to prove is a direct consequence of Proposition \ref{prop:drift_exponential_decay}, given the point-wise exponential decay of $\rho$. 
\end{proof}

We now conclude that using a time-dependent external electric field, we not only suppress the instability, but also ensure that the density perturbation and the self-consistently generated electric field decay superexponentially to zero. This, however, is an asymptotic result in time and indeed we see in the numerical simulations (see Section \ref{sec:numerical}) that initially the electric field is subject to a certain amount of oscillations. 

{It is worth noting that the pole elimination strategy merely eliminates the positive roots, and thus getting rid of the possibility of having exponential growth of the electric field. The superexponential decay is only obtained because the $h=0$ scenario coincides with the recovering of the free-streaming effect.}

\section{Application onto Two-Stream Instability and Bump-on-Tail Instability}\label{sec:numerical}

This section is devoted to the numerical validation of the proposed method for mitigating instabilities in plasma using an external electric field. The numerical implementation follows the method in~\cite{einkemmer2019performance,filbet2003comparison}. Specifically, we employed a semi-Lagrangian type algorithm utilizing cubic splines for spatial approximation to overcome the strict CFL condition. We present our numerical findings for both a two-stream instability problem and a bump-on-tail instability problem, demonstrating the effectiveness of our approach.

We first notice that both the two-stream and the bump-on-tail scenarios have the equilibrium state $\mu(v)$ in the form of a Gaussian mixture, so some computation is repeated. As a preparation, to unify the computation, we first prepare the calculation of ${\hat{U}}(t,k)$ in this general setting.

To start, we set \( W_{m,\sigma} = \mathcal{N}(m, \sigma) \) as a Gaussian distribution with mean \( m\) and a standard deviation of $\sigma$. Given a series of real number pairs $\{m_j, \sigma_j\}_{j=1}^N$, we consider an equilibrium that combines these Gaussian distributions:
\begin{equation*}
   \mu(v) = \sum_{j=1}^N \alpha_j W_{m_j,\sigma_j}(v),
\end{equation*}
where $\alpha_j$ are real number weights and $\sum_{j=1}^N \alpha_j=1$ for normalization. Our objective is to examine the form of \( \hat{U} \) in this scenario and derive an explicit construction for \( \hat{H} \) using \eqref{eq:h(x)=x_choice}.

Given the Gaussian form of \( W_{m,\sigma} \), we can explicitly compute its Fourier transform:
\begin{align*}
    \hat{W}_{m,\sigma}(k) &= \int_{-\infty}^{\infty} W_{m,\sigma}(v) e^{-ikv} \, dv = \int_{-\infty}^{\infty} \frac{1}{\sqrt{2\pi}\sigma} e^{-\frac{1}{2}\frac{(v - m)^2}{\sigma^2}} e^{-ikv} \, dv \\
    &= \frac{e^{-ikm}}{\sqrt{2\pi}} \int_{-\infty}^{\infty} e^{-\frac{1}{2}y^2 -ik\sigma y } \, dy\\
    & =\frac{e^{-ikm}}{\sqrt{2\pi}}  \int_{-\infty}^\infty e^{-\frac{1}{2}(y^2+2ik\sigma y-k^2\sigma^2)}\,e^{-\frac{1}{2}k^2\sigma^2}\,dy  \\
    &= e^{-ikm - \frac{1}{2}k^2\sigma^2}\,.
\end{align*}
Utilizing linearity, we have
\begin{equation*}
    \hat{U}(t, k) =  t \sum_{j=1}^N \alpha_j e^{-\frac{1}{2}k^2\sigma_j^2t^2} e^{-iktm_j}\,.
\end{equation*}
Following this derivation, and deploying~\eqref{eq:h(x)=x_choice}, we have, when the equilibrium presents the form of a linear combination of Gaussians, $H$ is explicit:
\begin{equation}
    \label{eq:H_hat_formula}
    \hat{H}(t,k) = \frac{1}{ik}\hat{S}(t,k) + \frac{\alpha}{ik}\left(\delta(t) + t e^{-\frac{1}{2}k^2\sigma^2t^2}\sum_{j=1}^N \alpha_j e^{-ikt\mu_j}\right),
\end{equation}
for $k\neq 0$ and \(\hat{H}(t,0)=0\). This formula provides a precise, closed-form solution for \(\hat{H}(t,k)\), which will be applied in the subsequent numerical experiments.

\subsection{Two-Stream Instability}\label{subsec:Two-Stream}

This subsection is dedicated to suppressing the two-stream instability. We restate the governing equation \eqref{eq:VP_system_pert} for the perturbation here:
\begin{equation*}
\left\{
    \begin{aligned}
        &\partial_t f(t,x,v) + v \cdot \nabla_x f(t,x,v) - \left[E(t,x) + H(t,x)\right] \cdot \nabla_v \left[f(t,x,v) + \mu(v)\right] = 0, \\
       &\Delta V(t,x) = -\rho(t,x), \\
       &E(t,x) = \nabla V(t,x),
    \end{aligned}
\right.
\end{equation*}
with the initial condition now specified as:
\[
f(0, x, v) = \varepsilon \cos(\beta x) \,\mu(v),
\]
and the equilibrium is set to be:
\[
\mu(v) = \frac{1}{2\sqrt{2\pi}} \left( \exp\left(-\frac{(v - \bar{v})^2}{2}\right) + \exp\left(-\frac{(v + \bar{v})^2}{2}\right) \right).
\]
Numerically we choose parameters \( \beta = 0.2 \) and \( \bar{v} = 2.4 \), various of values for the parameter \( \varepsilon \). The numerical domain is set to be $(x, v) \in \left[0, \frac{2\pi}{\beta}\right] \times [-6, 6]$, and discretized using 128 grid points in both the spatial and velocity dimensions, and the time step size \( \Delta t = 0.1 \) is chosen. Numerical experiments suggests this choice of discretization is sufficiently accurate.

Using numerical simulation, drawing from prediction from Section \ref{sec:control}, we study four scenarios:
\begin{itemize}
    \item System A:
    This is the original system without control, i.e.~we set $H=0$;
    \item System B, with $h(x) = 0$: This corresponds to setting $H=-E[f_\rmf]$ and completely removes the self-generated electric field. It is the field neutralization situation and is expected to recover free-streaming;
    \item System C, with $h(x) = \alpha x$: This corresponds to removing the roots in the linear stability analysis, and is expected to be effective only in the linear regime. 
    \item System D, with $h(x) = \alpha x$ before $t_0$ and transition into $h(x) = 0$ after $t_0$: This is the combined method to utilize $\alpha x$ to strike down singularity and use free-streaming to maintain damping. Numerically we set $t_0=10$.
\end{itemize}
Numerically, we are to validate these predictions. We also present some numerical findings that are poorly understood.

\noindent\textbf{Numerical validation:}
We first numerically evaluate the suppressing features of different types of $h$, to validate the predictions derived from the theory. Throughout this section, we set \(\varepsilon = 0.005\) in the initial perturbation, and choose \(\alpha = 0.05\).

\begin{figure}[ht]
    \centering
    \begin{minipage}{0.48\textwidth}
        \centering
        \includegraphics[width=\textwidth]{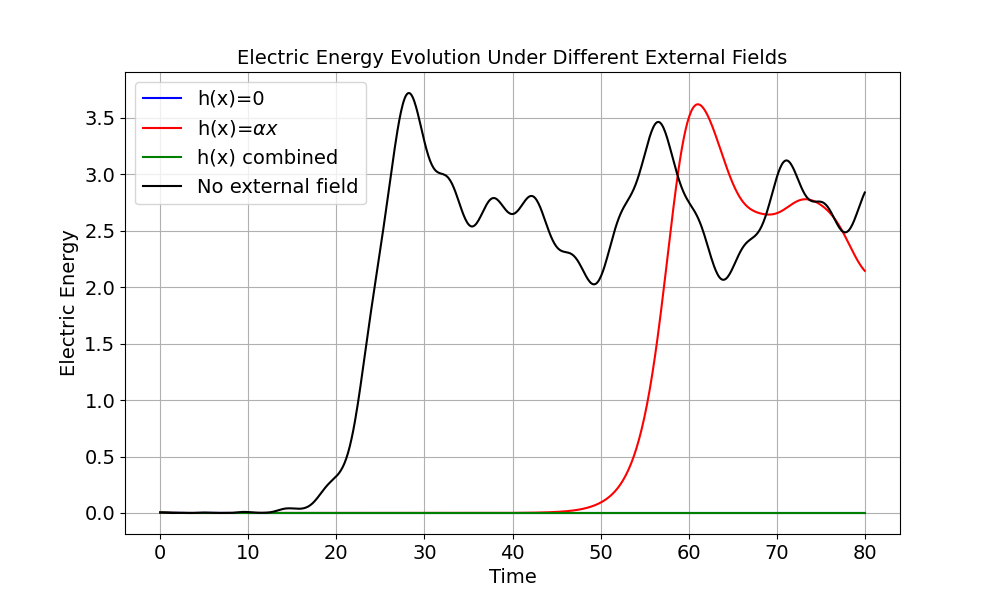}
        \caption*{(a) Electric energy evolution under different external fields for two stream case: comparison over the interval \([0,80]\)}
    \end{minipage}%
    \hspace{0.03\textwidth}
    \begin{minipage}{0.48\textwidth}
        \centering
        \includegraphics[width=\textwidth]{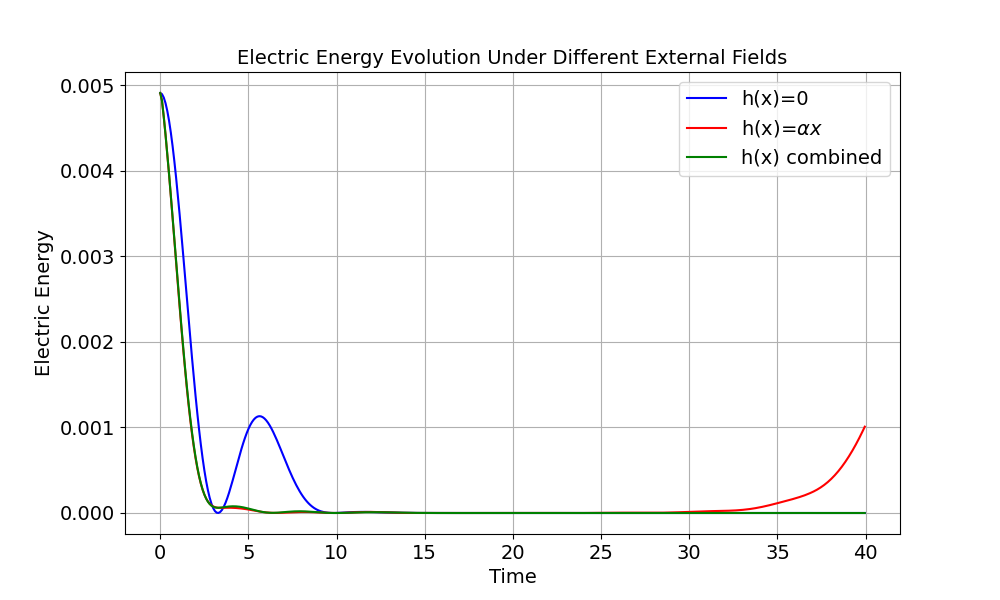}
        \caption*{(b) Electric energy evolution under different external fields for two stream case: comparison over the interval $[0,40]$}
    \end{minipage}
  \caption{Comparison of electric energy evolution under different external fields {(System A to D)} for the two-stream case with $\alpha = 0.05$ in {System C}. }
    \label{fig:electric_energy_evolution} 
\end{figure}

We first plot the electric energy as a function of time for all four $h$ choices. In the time frame of $[0,80]$, as shown in Figure~\ref{fig:electric_energy_evolution}, part (a), for System A, the electric energy demonstrates a characteristic exponential increase (black curve). This pattern continues until the electric field reaches a level where significant nonlinear effects become dominant (around $t=30$), and the field strength saturates. The red curve plots the electric energy of System C. At the initial time when the system is still in the linearized regime, the external field successfully suppresses the electric energy growth. At around $t=40$, linear analysis loses its accuracy. Once the system is out of the linearized regime, merely removing the roots is no longer effective. The electric energy started picking up again. On the contrary, both System B and D show suppressed electric energy throughout the entire evolution.

In Figure~\ref{fig:electric_energy_evolution}, part (b), we zoom in to the time frame of $[0,40]$. In the beginning period ($t\in[0,10]$), the free-streaming case, System B with $h(x) = 0$ still presents some oscillations, while System C shows a nice (almost) monotone decay of the electric energy. As time propagates to around $t=35$, the nonlinear effects pick up, making the strategy in System C invalid, so the red curve starts curving up. System D combines two strategies, and switch from $h(x)=\alpha x$ to $h(x) = 0$ at time $t=10$. Numerically this seems to have given us the optimal control.

\begin{figure}[htp]
    \centering
    \begin{minipage}{6.5in}
        \centering
        \includegraphics[width=0.27\textwidth]{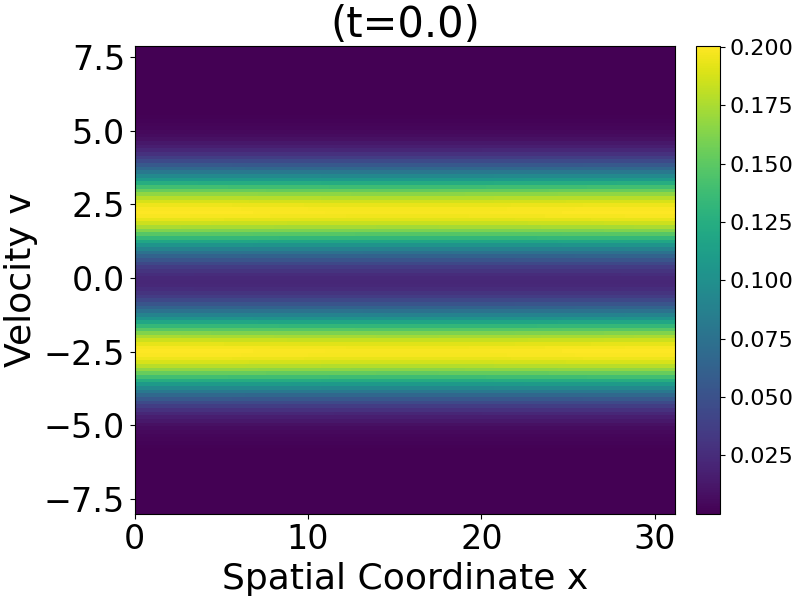}
        \includegraphics[width=0.27\textwidth]{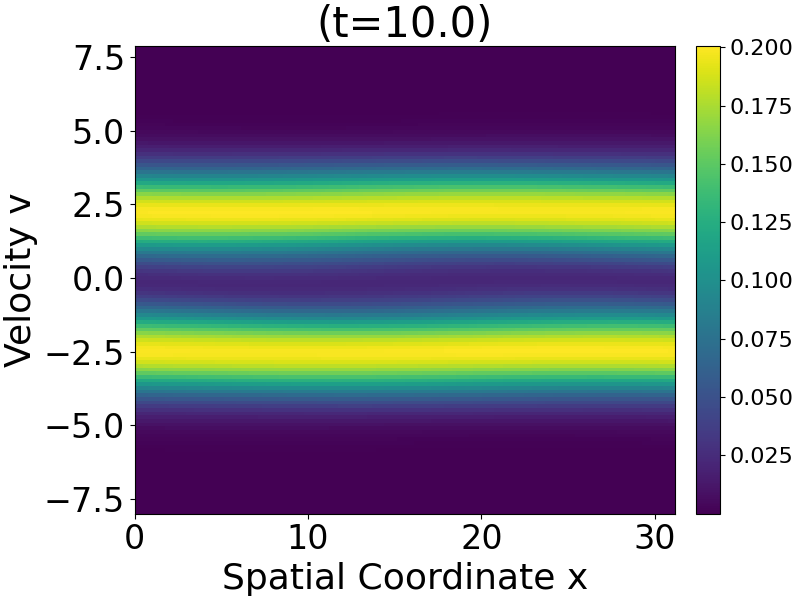}
        \includegraphics[width=0.27\textwidth]{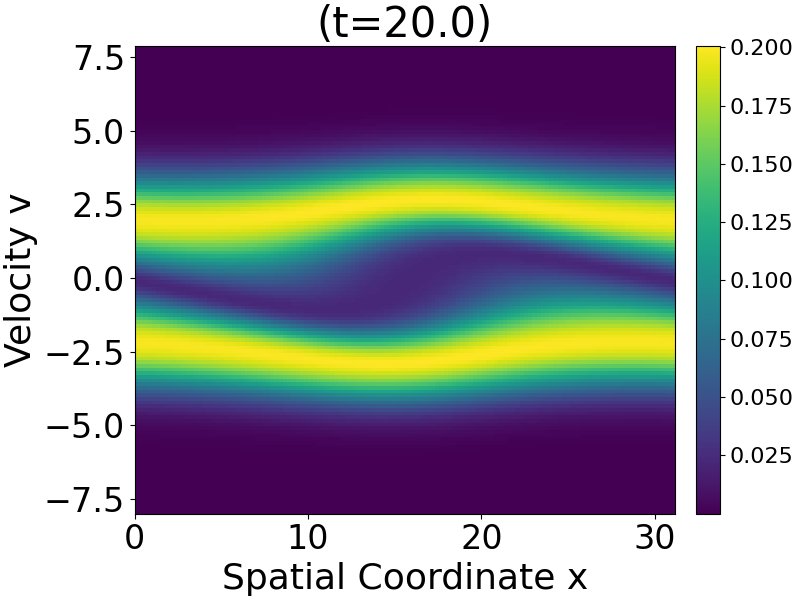}
        
        \includegraphics[width=0.27\textwidth]{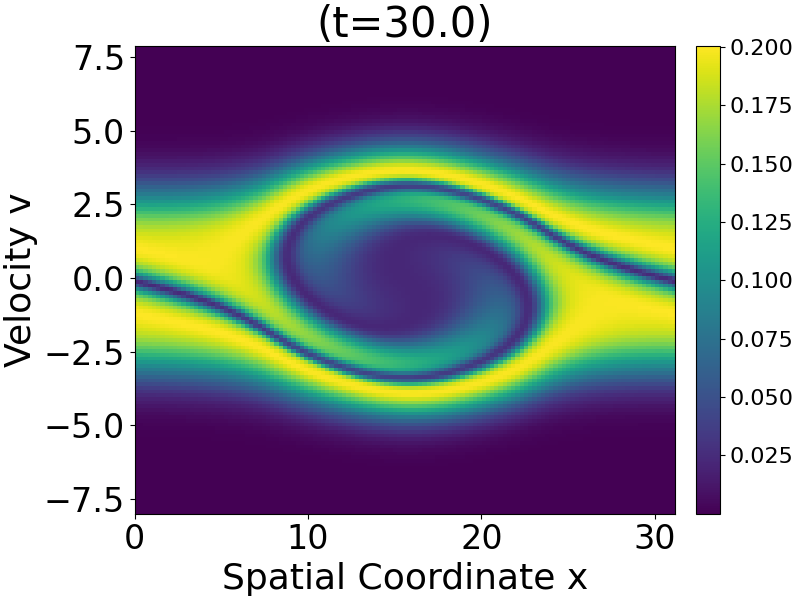}
        \hspace{0.1in}\includegraphics[width=0.27\textwidth]{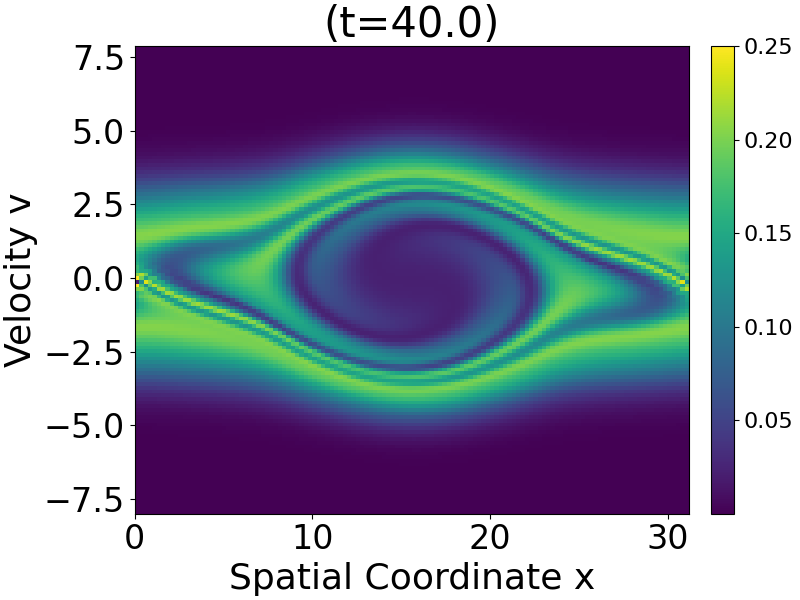}
        \includegraphics[width=0.27\textwidth]{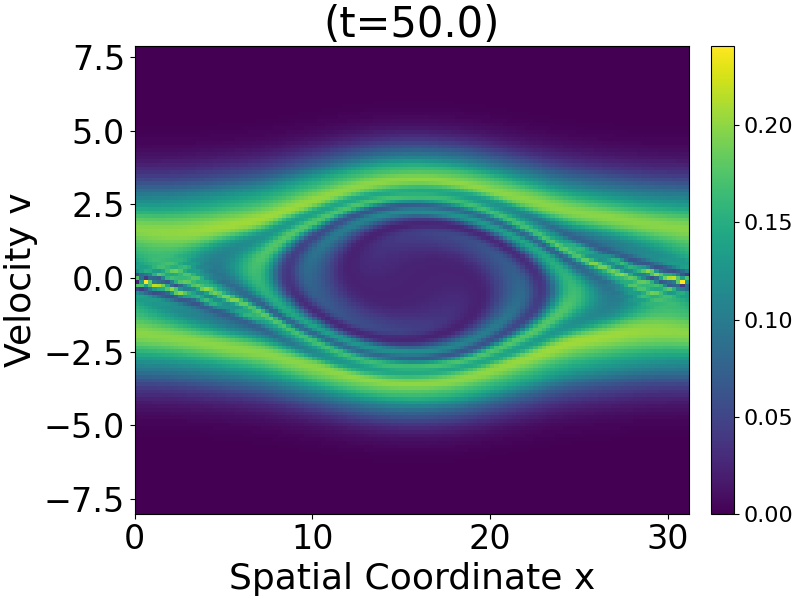}
        
        \includegraphics[width=0.27\textwidth]{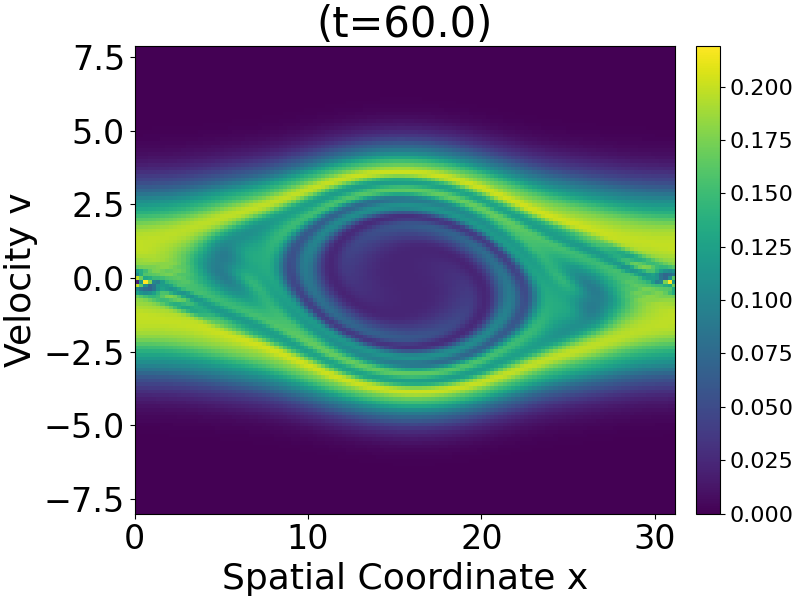}
        \includegraphics[width=0.27\textwidth]{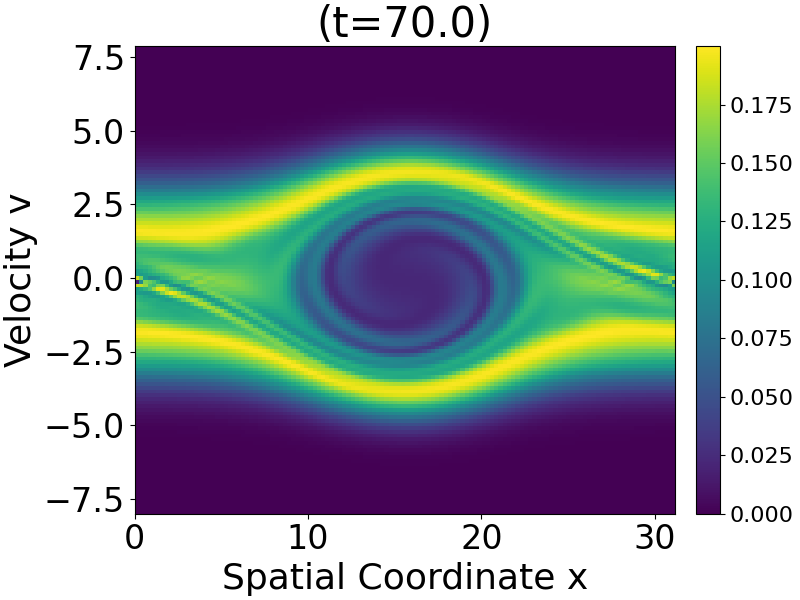}
        \includegraphics[width=0.27\textwidth]{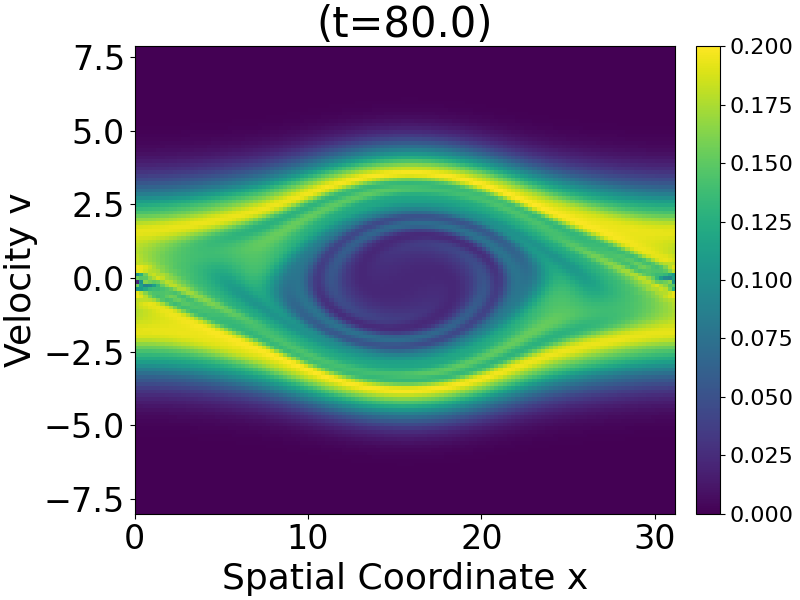}
        
        \caption*{(a) Snapshots for System A}
    \end{minipage}
    
    \vspace{1cm} 
    
    \begin{minipage}{6.5in}
        \centering
        \includegraphics[width=0.27\textwidth]{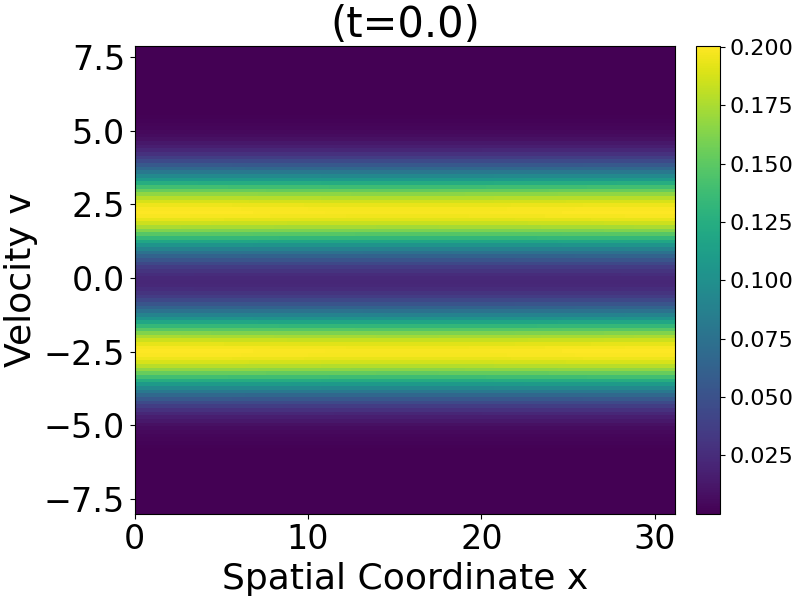}
        \includegraphics[width=0.27\textwidth]{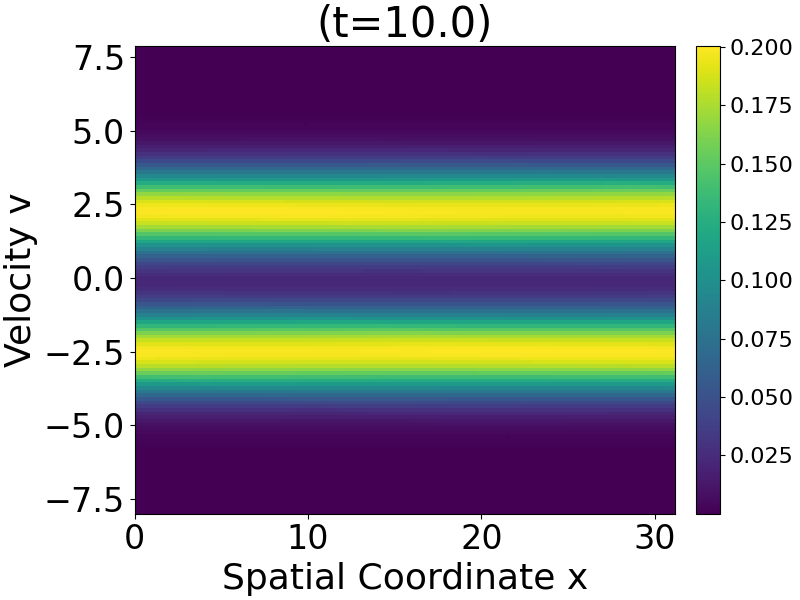}
        \includegraphics[width=0.27\textwidth]{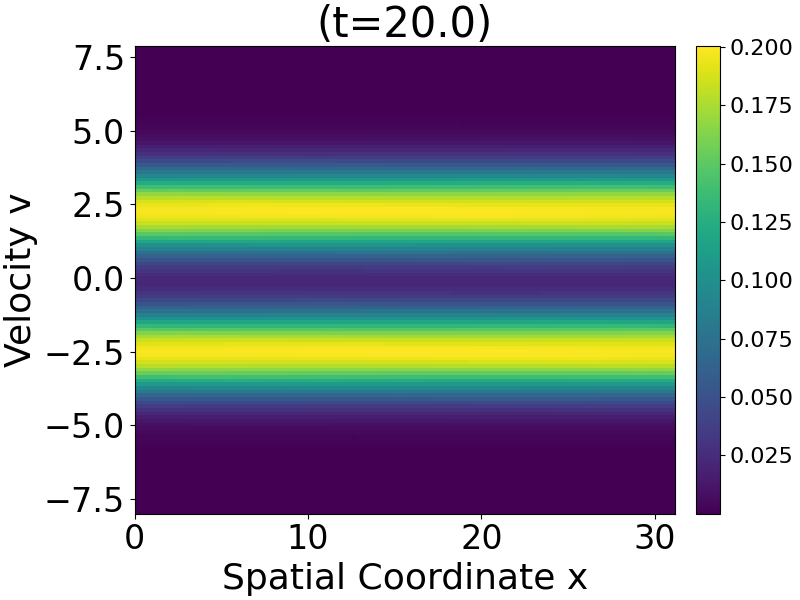}
        \includegraphics[width=0.27\textwidth]{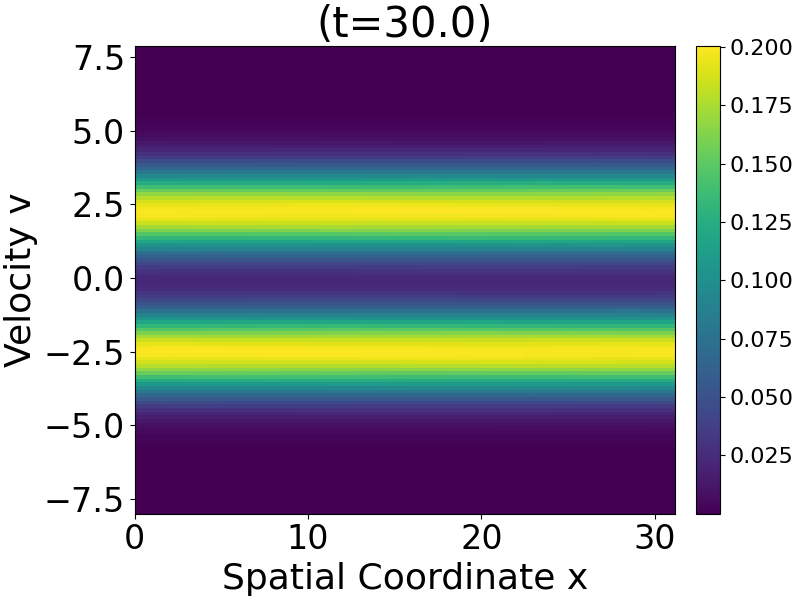}
        \includegraphics[width=0.27\textwidth]{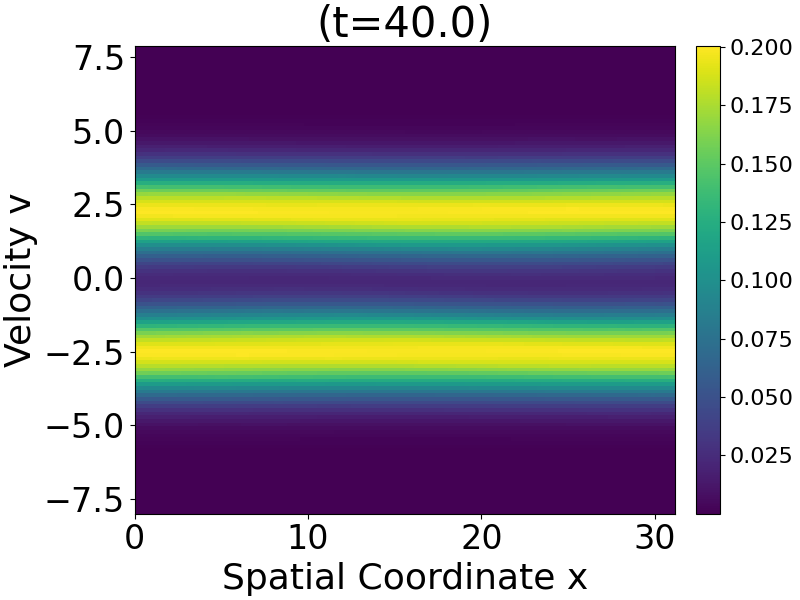}
        \includegraphics[width=0.27\textwidth]{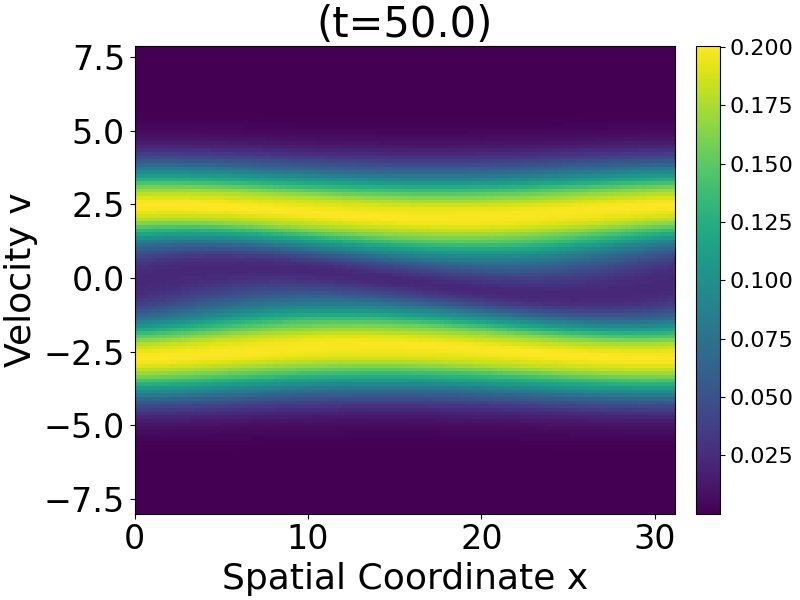}
        \includegraphics[width=0.27\textwidth]{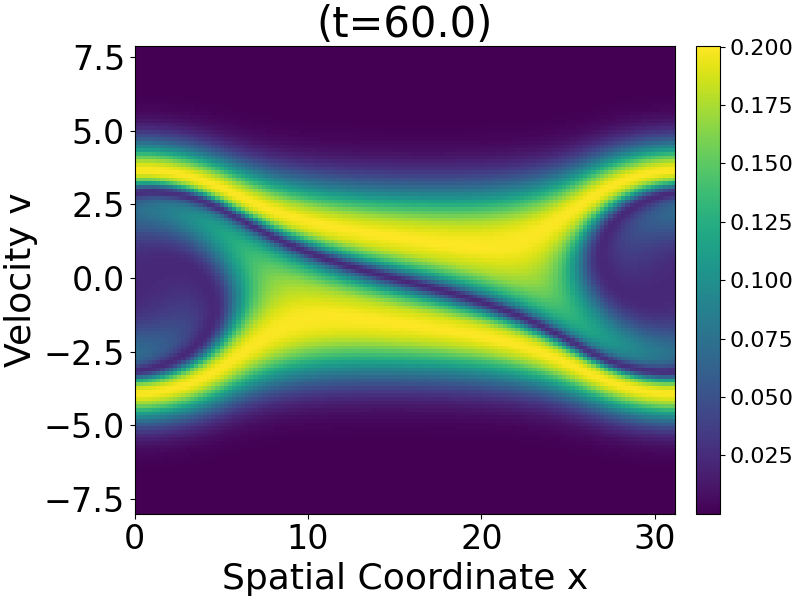}
        \includegraphics[width=0.27\textwidth]{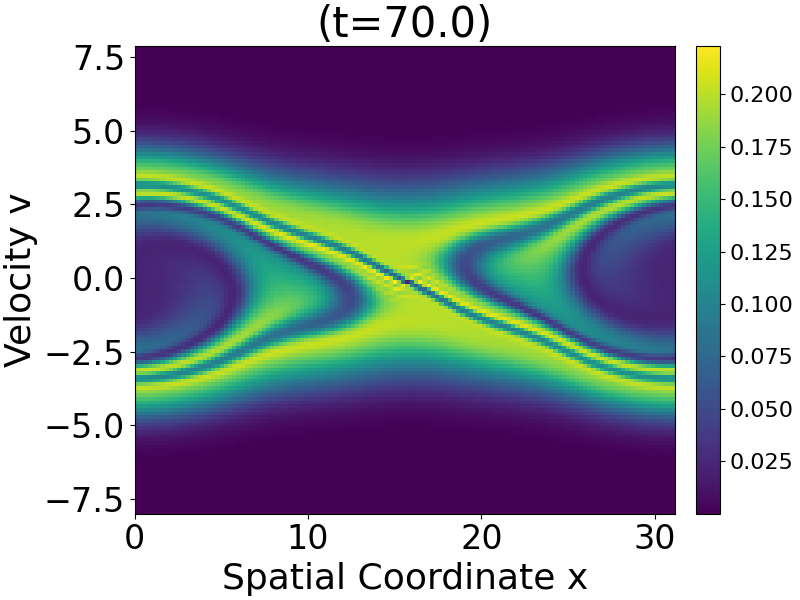}
        \includegraphics[width=0.27\textwidth]{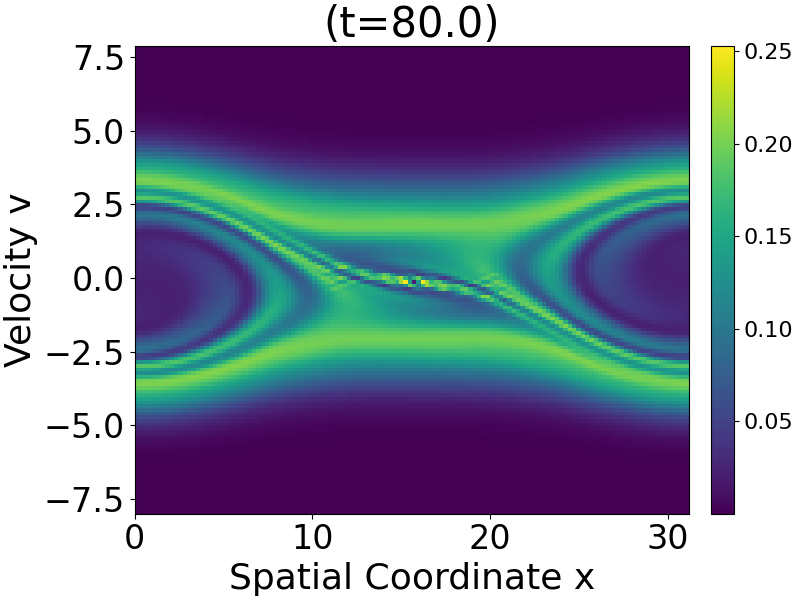}
        
        \caption*{(b) Snapshots for System C with $h(x) = 0.05 x$}
    \end{minipage}
    
    \caption{Comparative snapshots under different conditions for two stream case.}
    \label{fig:comparative_snapshots}
\end{figure}

With a better understanding of electric energy, we now examine the evolution of the entire plasma distribution over the phase space \((x,v)\).  In Figure \ref{fig:comparative_snapshots}, we present a series of snapshots for System A and System C to illustrate the delayed onset of chaotic behavior. Specifically, when control is absent in System A, the plasma starts swirling and exhibits chaotic behavior between \(t=10\) and \(t=20\). In contrast, System C prevents this chaotic behavior during the same time frame. However, once the validity of the linearized regime is lost, chaotic behavior re-emerges between \(t=40\) and \(t=50\).  Thus, we observe a delayed onset of chaos rather than its complete elimination. This observation is consistent with our findings on the growth of electric energy, as shown in Figure \ref{fig:electric_energy_evolution}, part (a). {Meanwhile, we observe that between $t = 60$ and $t = 80$, systems A and C exhibit different phases in the $x$-$v$ space, which can be attributed to the nonlinear interaction between the plasma and the external electric field.
}

\begin{figure}
    \centering
    \includegraphics[width=7in]{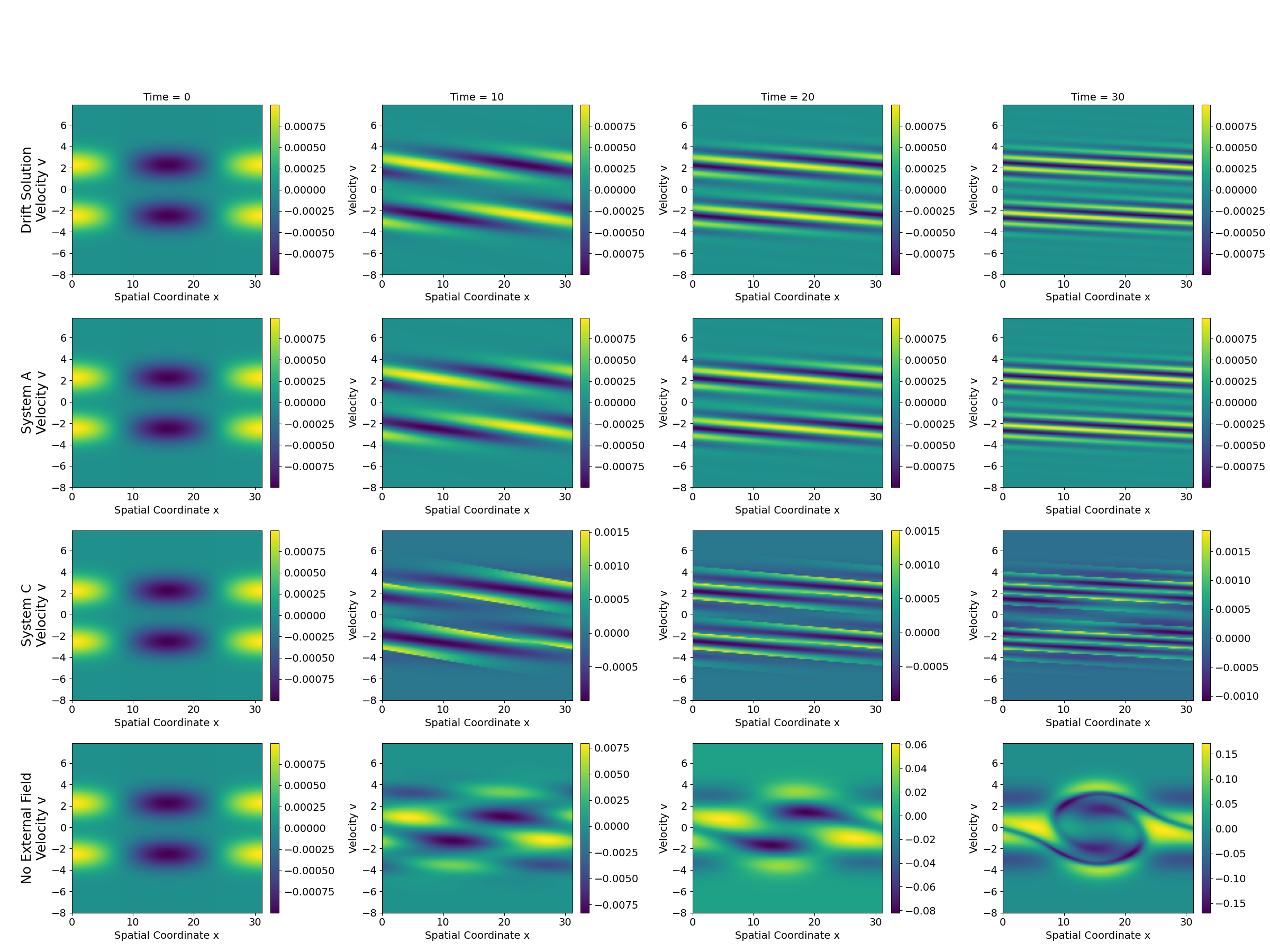}
    \caption{{Comparative evolution of perturbation in various scenarios. The four rows illustrate solutions to different equations. The first row shows the solution to the free-streaming equation~\eqref{eq:drift_equation_mod}. The second and third rows display the solutions to the VP system~\eqref{eq:VP_system_pert} with an external field, corresponding to System B and System C, respectively. Finally, the fourth row presents the solution to the original VP system with \( H = 0 \), corresponding to System A.
}}
    \label{fig:filament}
\end{figure}

To further demonstrate the effectiveness of the electric field neutralization approach described in Section \ref{sec:negating_electric_field} for system B, and to confirm the recovery of free-streaming behavior, we conducted a simulation focusing on the evolution of the perturbation component in a two-stream scenario. Figure \ref{fig:filament} presents a color-coded profile of the distribution function \( f \) in phase space at four distinct times: \( t = 0, 10, 20, 30 \). {The figure consists of four rows, each corresponding to a different scenario: the free-streaming equation~\eqref{eq:drift_equation_mod}, System B, System C, and the original VP system with \( H = 0 \) (System A). The third row (system C) is included for comparison with system B. The first two rows, which display identical filamentation leading to a homogenized electric field, validate Proposition \ref{prop:drift_exponential_decay} by demonstrating consistent behavior. In contrast, system C also exhibits filamentation, but its structure differs from the first two rows, as seen in the graph. Finally, the absence of an external field in the fourth row results in a swirling phase space profile, which corresponds to an exponential increase in electric energy.
}

\vspace{0.2cm}
\noindent\textbf{Numerical findings:}
With numerical evidence validating theory, we now proceed to explore scenarios that are yet beyond theory. In particular, we investigate numerically the choice of \( \alpha \) in System C that hopefully optimally suppress instability.

Based on equation \eqref{eq:rho_h(x)=x_physical}, we anticipate that the density function \(\rho\) will exhibit a pattern similar to \(S + \alpha U\). Given that \( E = -\nabla\Delta^{-1}\rho \), minimizing the \( L_2 \) norm of \( E \) mathematically corresponds to minimizing the \( H^{-1} \) norm of \(\rho\). Intuitively, a flatter \(\rho\) will result in a lower electric field. In other words, by choosing a suitable value for \(\alpha\), if we can make \(S + \alpha U\) flatter than \(S\) itself, we will potentially achieve a smaller electric field compared to System B, which corresponds to \(\alpha = 0\).

\begin{figure}[htb]
    \centering
    \includegraphics[width=0.48\textwidth]{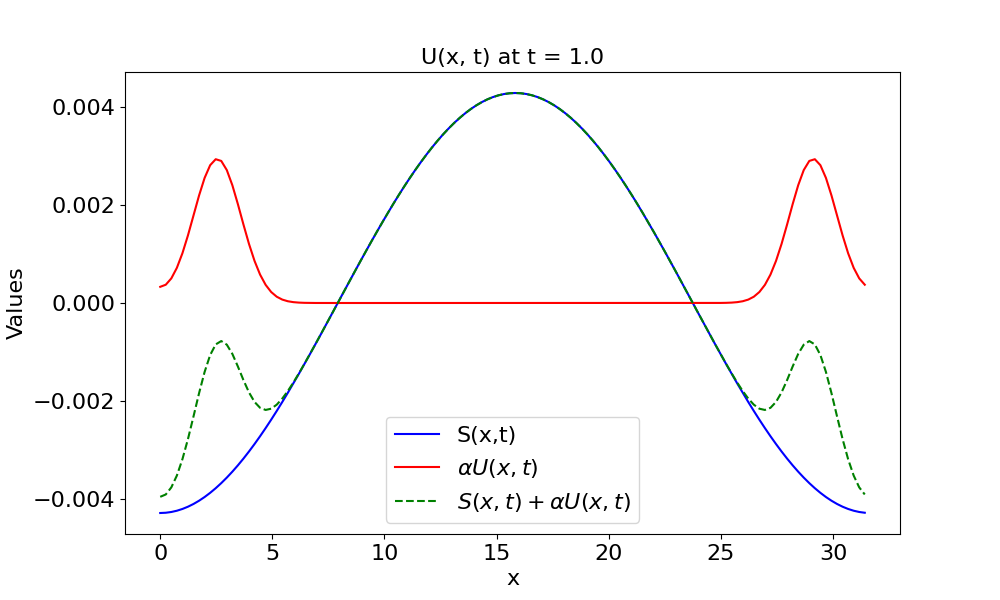}
    \includegraphics[width=0.48\textwidth]{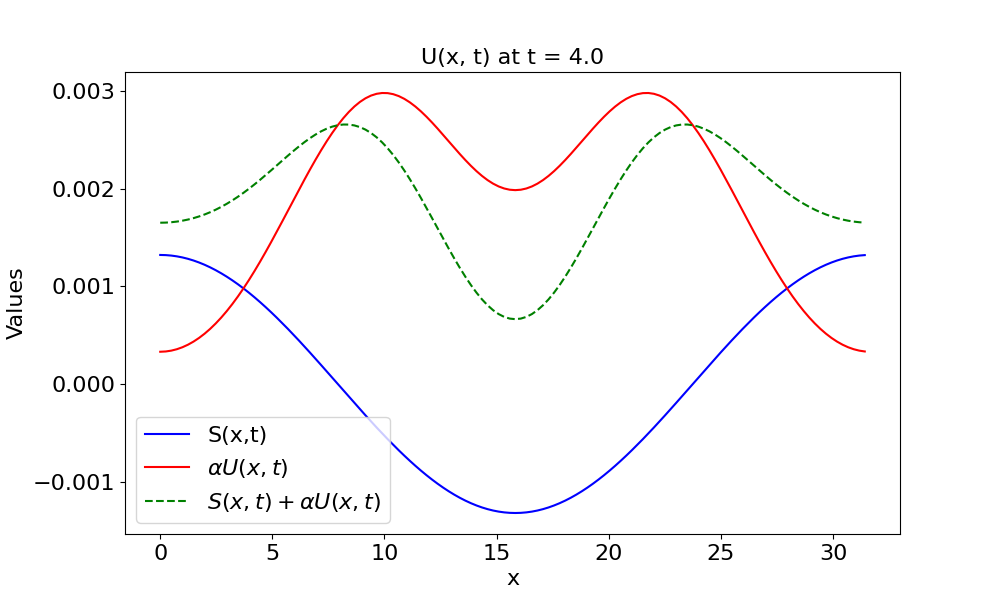}

     \vspace{0.5cm} 

    
    \includegraphics[width=0.48\textwidth]{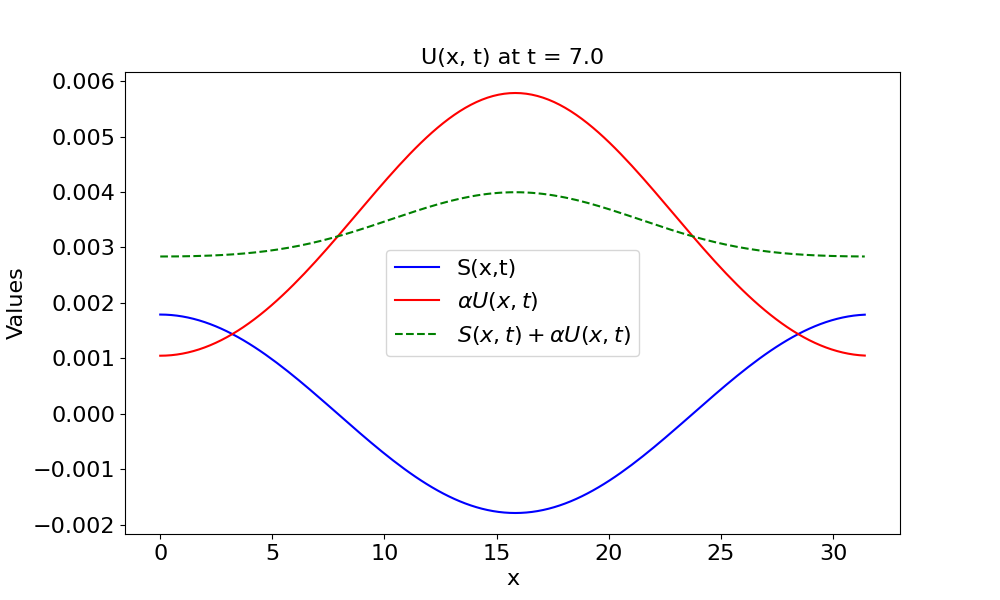}
    \includegraphics[width=0.48\textwidth]{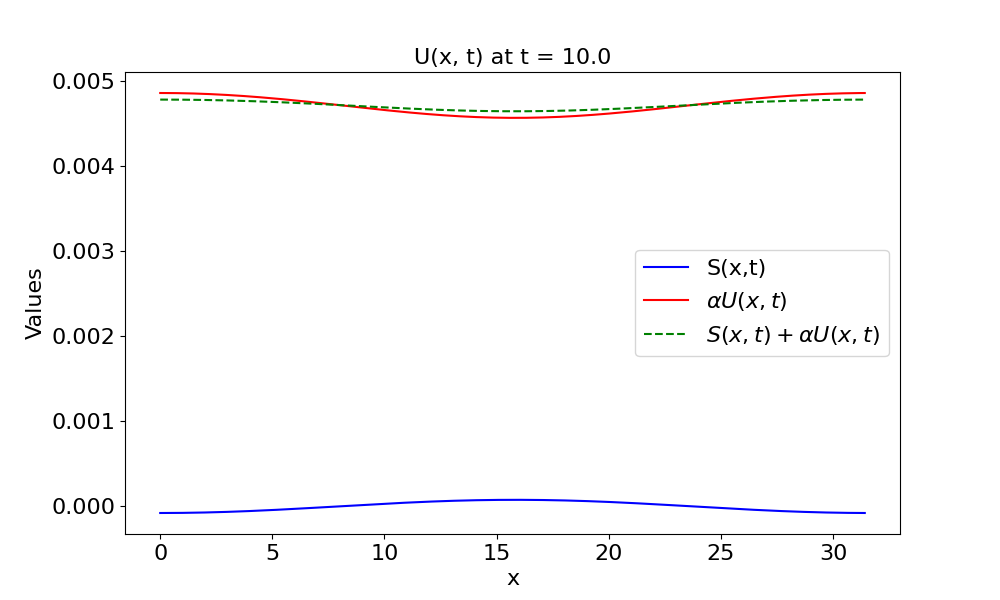}
    
    \caption{Graph of $S(t,x), \alpha U(t,x)$ and $S(t,x)+\alpha U(t,x)$ for two stream case at different moments}
    \label{fig:S+alphaU}
\end{figure}

With this in mind, we plot the evolution of \(S(t,x)\), \(U(t,x)\), and \(S + \alpha U\) with \(\alpha = 0.05\), which is our choice for the experiments here. In Figure \ref{fig:S+alphaU}, we show \(S(t,x)\) in blue, \(\alpha U(t,x)\) in red, and their cumulative impact in green. We can see that at the moments presented here, the red curve and the blue curve cancel each other out, resulting in a flatter green curve compared to the blue curve. This justifies the validity of choosing \(\alpha = 0.05\) based on our discussion.

To be more specific, the plot shows that \( U(t,x) \) initially presents a two-stream-like distribution with two distinct peaks. Over time, these peaks move towards each other and merge. At $t=5$ after they completely merge, the contribution from $U$ seems to cancel out that in $S$, leaving the combination (green line) roughly flat. As a consequence, the electric energy is small.

\begin{figure}[ht]
    \centering
    \begin{minipage}{0.48\textwidth}
        \centering
        \includegraphics[width=\textwidth]{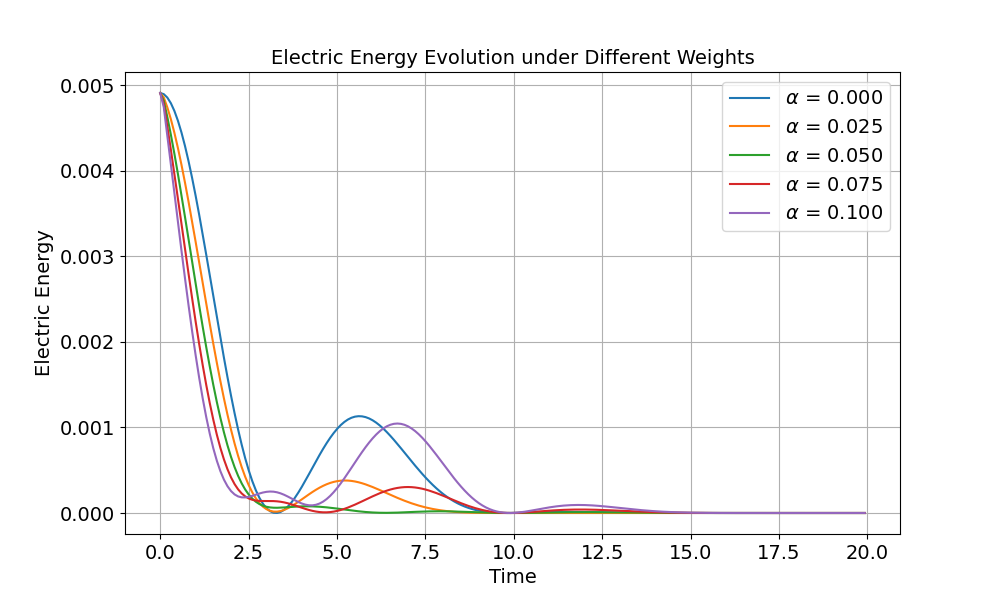}
        \caption*{(a) Comparison over the time interval \([0,20]\) with weights $\alpha$ within $[0,0.1]$.}
    \end{minipage}%
    \hspace{0.03\textwidth} 
    \begin{minipage}{0.48\textwidth}
        \centering
        \includegraphics[width=\textwidth]{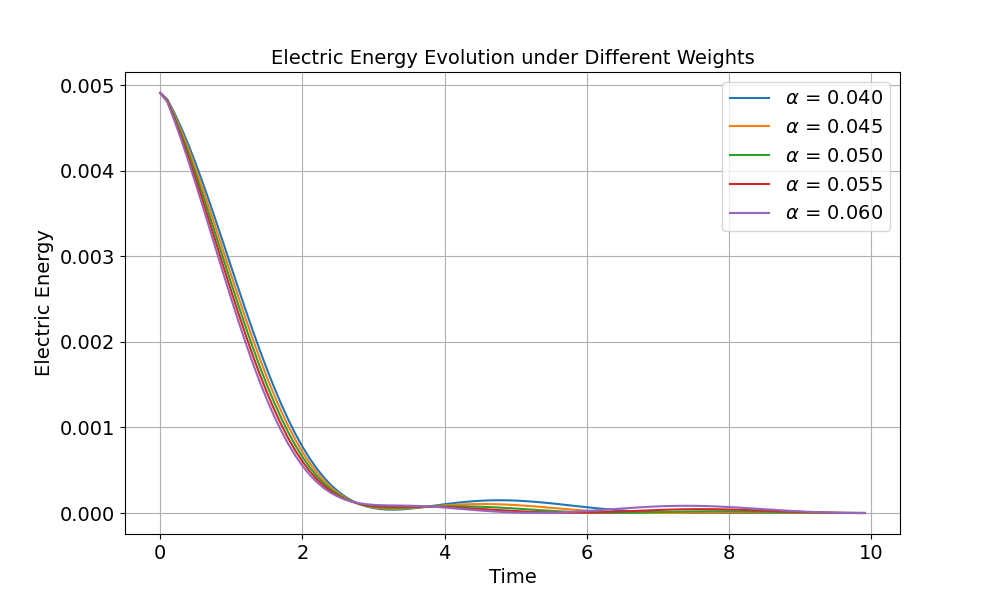}
        \caption*{(b) Comparison over the time interval \([0,10]\) with weights $\alpha$ within $[0.04,0.06]$}
    \end{minipage}
   \caption{{Electric energy evolution for different weights used in System C when applied to the two-stream instability.}}

    \label{fig:electric_energy_different_weights_ts}
\end{figure}

We also present a comparison of electric energy evolution under different values of \(\alpha\). In Figure \ref{fig:electric_energy_different_weights_ts}, we plot the evolution of electric energy over time for various weights. Since the electric energy becomes very small after \(t=20\), as shown in the figure, we focus on the early stage of the evolution rather than the entire time span \([0, 80]\). From Figure \ref{fig:electric_energy_different_weights_ts}, part (a), we observe that as the value of \(\alpha\) increases from 0 to 0.1, the electric energy initially decreases and then increases again, with \(\alpha = 0.05\) performing the best among the five different values tested. To identify a more accurate optimal choice of \(\alpha\), we zoom in to examine \(\alpha\) values from 0.04 to 0.06 over time span $[0,10]$ and compare their performance. From Figure \ref{fig:electric_energy_different_weights_ts}, part (b), we see that all these choices of weights effectively suppress the instability, with each value performing better or worse than the others at different times. Overall, all values in this range provide reasonable control, and \(\alpha = 0.05\) appears to be a well-balanced choice.

\subsection{Bump-on-Tail Instability}
In this subsection, our focus shifts to the bump-on-tail instability. Similar to the two-stream instability, we will conduct experiments to compare the behavior of plasma under various types of external control. We will consider system \eqref{eq:VP_system_pert} and define the four systems in the same manner as described in Section \ref{subsec:Two-Stream}. Unlike the previous subsection, we will perform the same experiments as above but will not differentiate between numerical validation and numerical findings.

The equilibrium state for this scenario is defined by the distribution:
\begin{equation*}
    \mu(v) = \frac{9}{10\sqrt{2\pi}}  \exp\left(-\frac{(v - \bar{v}_1)^2}{2}\right) + \frac{\sqrt{2}}{10\sqrt{\pi}}\exp\left(-2(v - \bar{v}_2)^2\right),
\end{equation*}
and the initial perturbation is set as:
\begin{equation*}
    f(0,x,v)=  \frac{\sqrt{2}\varepsilon}{10\sqrt{\pi}}\exp\left(-2(v - \bar{v}_2)^2\right)\cos(\beta x),
\end{equation*}
where \( (x, v) \) lies in the range \( \left[0, \frac{2\pi}{\beta}\right] \times [-9, 9] \), with the parameters \( \bar{v}_1 = -3 \), \( \bar{v}_2 = 4.5 \), \( \varepsilon = 0.05 \), and \( \beta = 0.1 \). The time interval is set to be \([0, 60]\) to cover the full range of dynamics. The parameter \(\alpha\) is selected to be \(-0.01\) for the reason discussed below.

\begin{figure}[ht]
    \centering
    \begin{minipage}{0.48\textwidth}
        \centering
        \includegraphics[width=\textwidth]{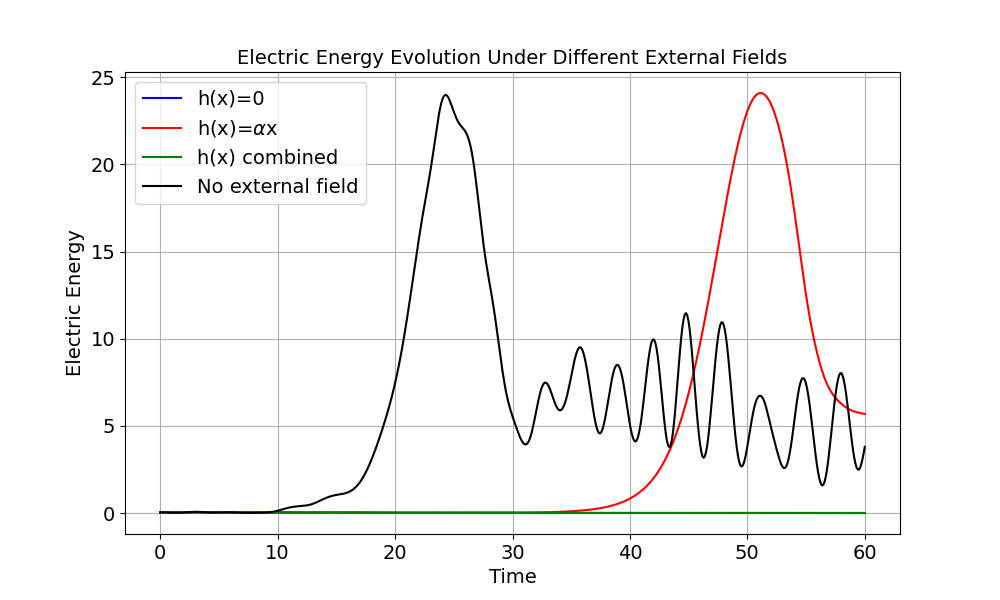}
        \caption*{(a) Electric Energy Evolution Under Different External Fields for Bump on Tail Case: Comparison over the Time Interval \([0,60]\)}
    \end{minipage}%
    \hspace{0.03\textwidth} 
    \begin{minipage}{0.48\textwidth}
        \centering
        \includegraphics[width=\textwidth]{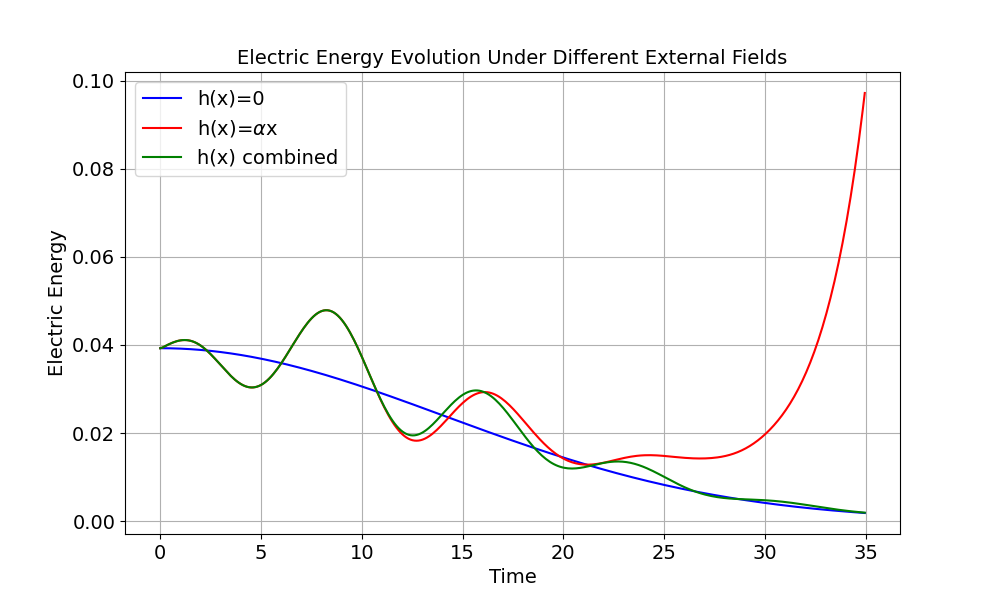}
        \caption*{(b) Electric Energy Evolution Under Different External Fields for Bump on Tail Case: Comparison over the Time Interval \([0,35]\)}
    \end{minipage}
    \caption{Comparative analysis of electric energy evolution in bump-on-tail instability under different external field strategies {(System A to D)} with $\alpha = -0.01$ {for System C}.}
    \label{fig:electric_energy_comparative_bot}
\end{figure}

Figure \ref{fig:electric_energy_comparative_bot} showcases the evolution of electric energy. As seen in part (a), System A, which has no external control, experiences a rapid increase in electric energy. System C initially suppresses the electric energy, but once nonlinear effects become significant (around \(t=30\)), the exponential increase resumes. In contrast, Systems B and D exhibit markedly different behavior, maintaining a much lower energy profile throughout the observed time interval. The zoomed-in version in Figure \ref{fig:electric_energy_comparative_bot}, part (b), allows us to observe the detailed behavior of Systems B, C, and D before the instability occurs. Unlike the two-stream case, System D does not consistently demonstrate more optimal behavior compared to System B. Instead, the difference in electric energy between Systems B and D oscillates, with System D sometimes suppressing the electric energy more effectively than System B and vice versa.

\begin{figure}[htp]
    \centering
    \begin{minipage}{6.5in}
        \centering
        \includegraphics[width=0.25\textwidth]{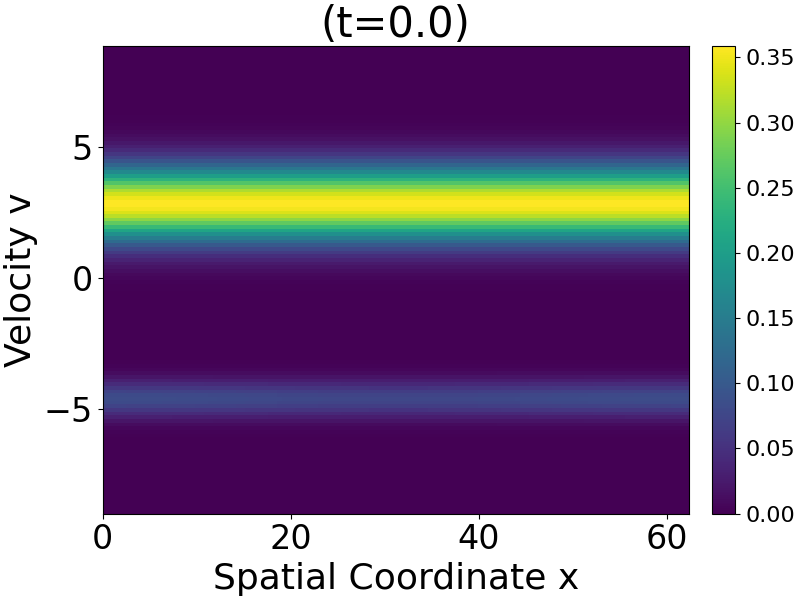}
        \includegraphics[width=0.25\textwidth]{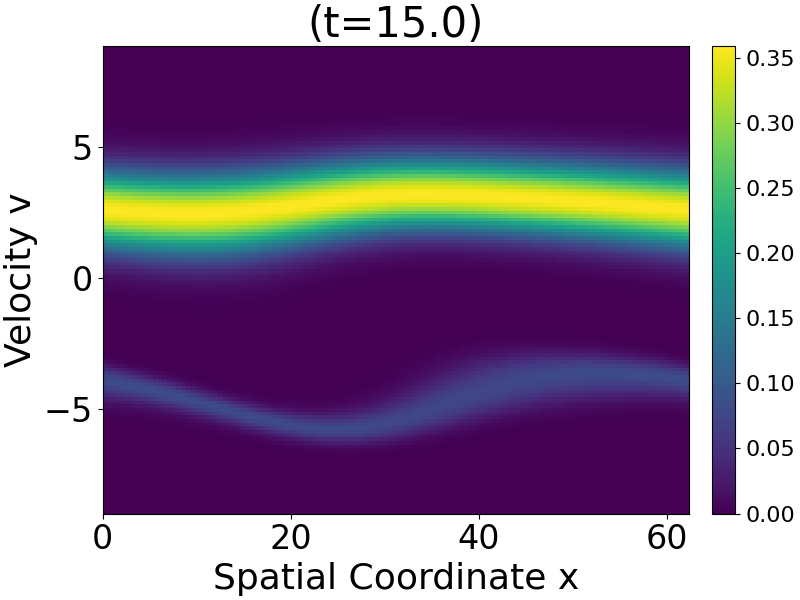}
        \includegraphics[width=0.25\textwidth]{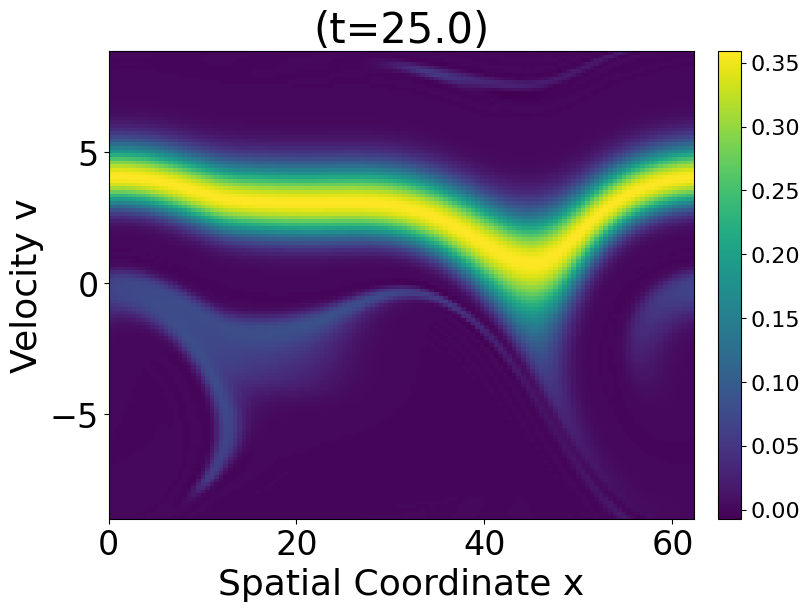}
        
        \vspace{0.5cm} 

        \includegraphics[width=0.25\textwidth]{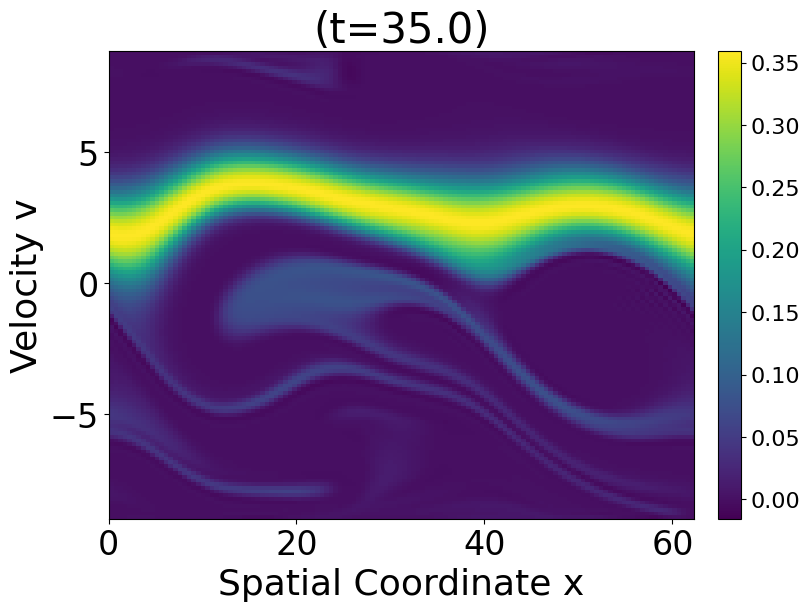}
        \includegraphics[width=0.25\textwidth]{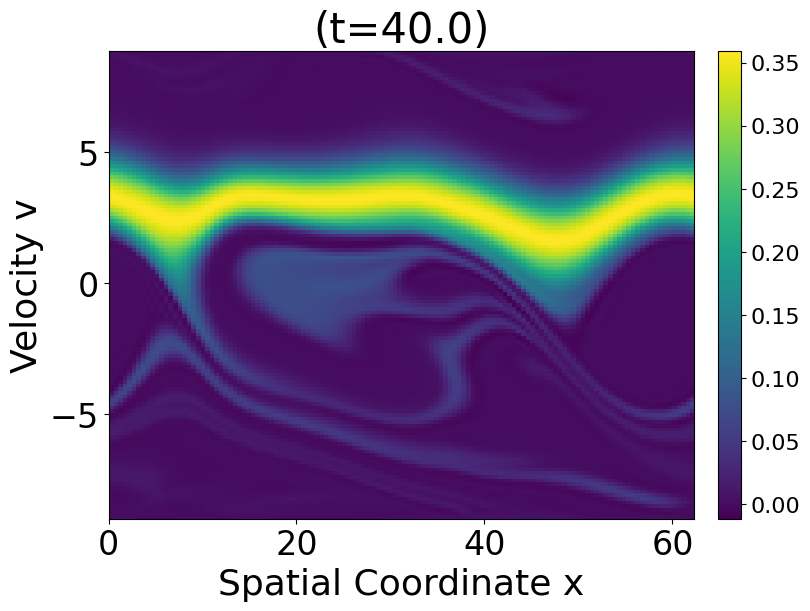}
        \includegraphics[width=0.25\textwidth]{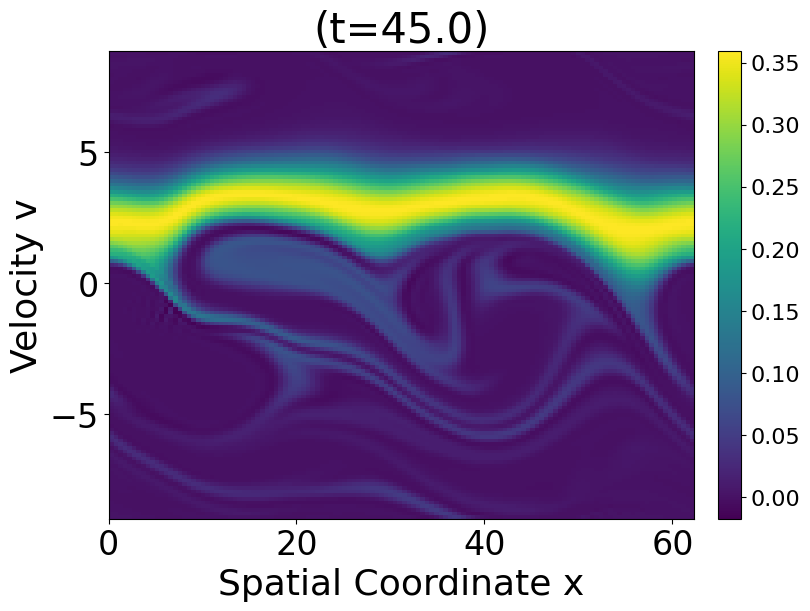}

        \vspace{0.5cm} 
        
        \includegraphics[width=0.25\textwidth]{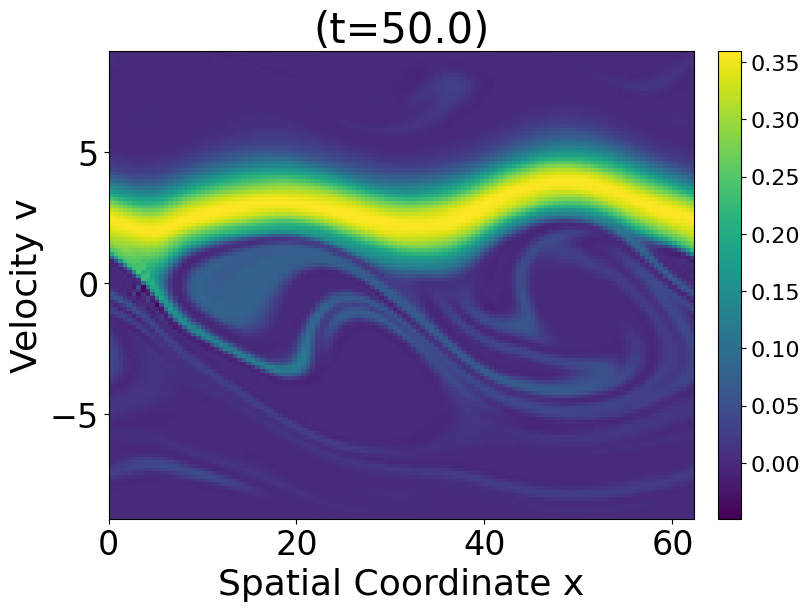}
        \includegraphics[width=0.25\textwidth]{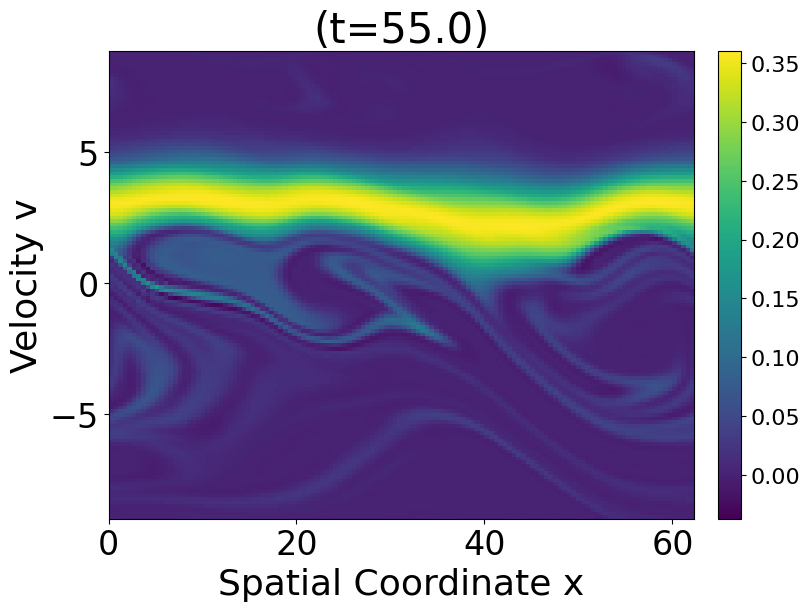}
        \includegraphics[width=0.25\textwidth]{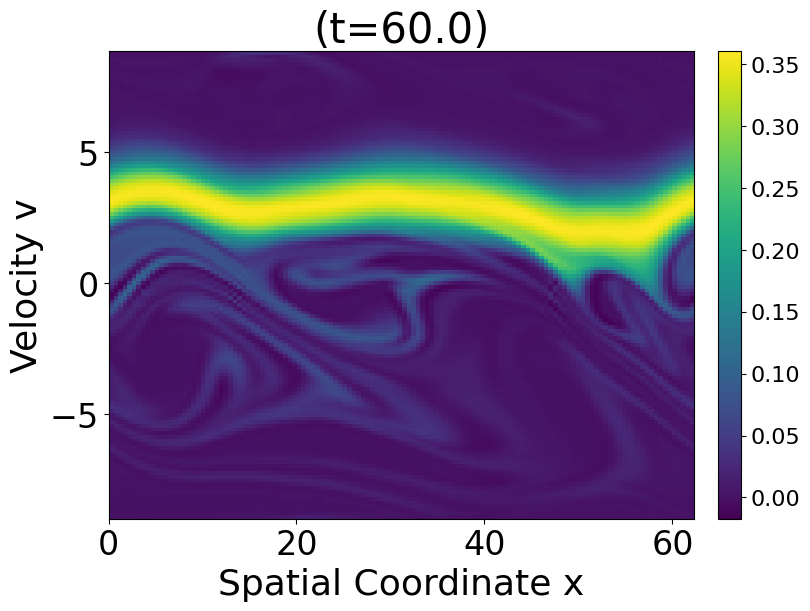}
        
        \caption*{(a) Snapshots for System A}
    \end{minipage}
    
    \vspace{1cm} 
    
    \begin{minipage}{6.5in}
        \centering
        \includegraphics[width=0.25\textwidth]{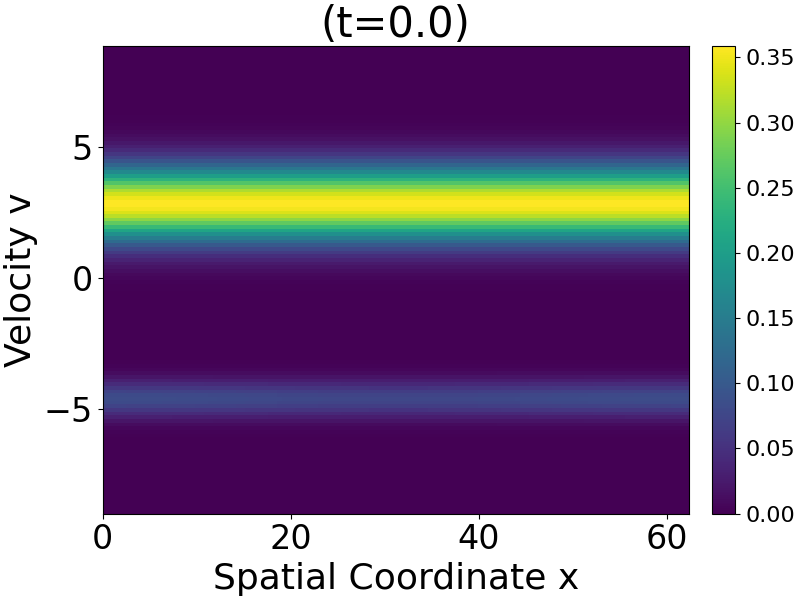}
        \includegraphics[width=0.25\textwidth]{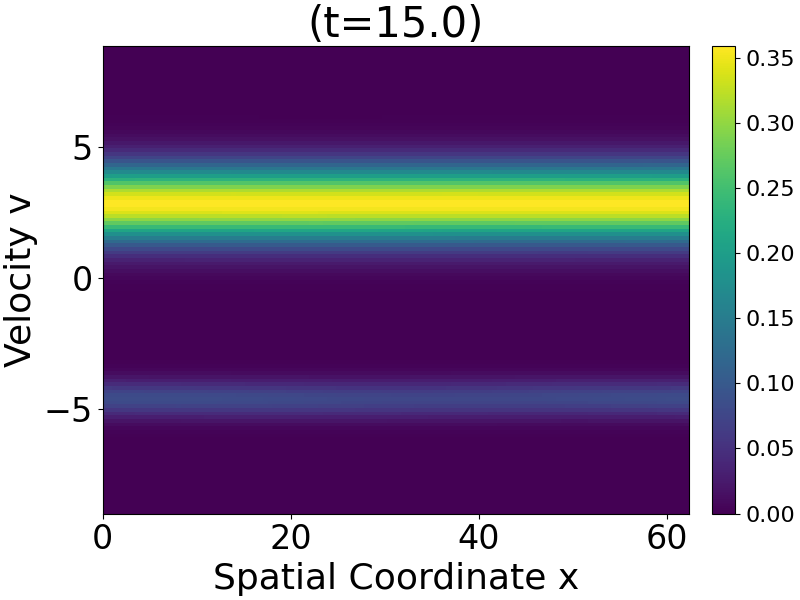}
        \includegraphics[width=0.25\textwidth]{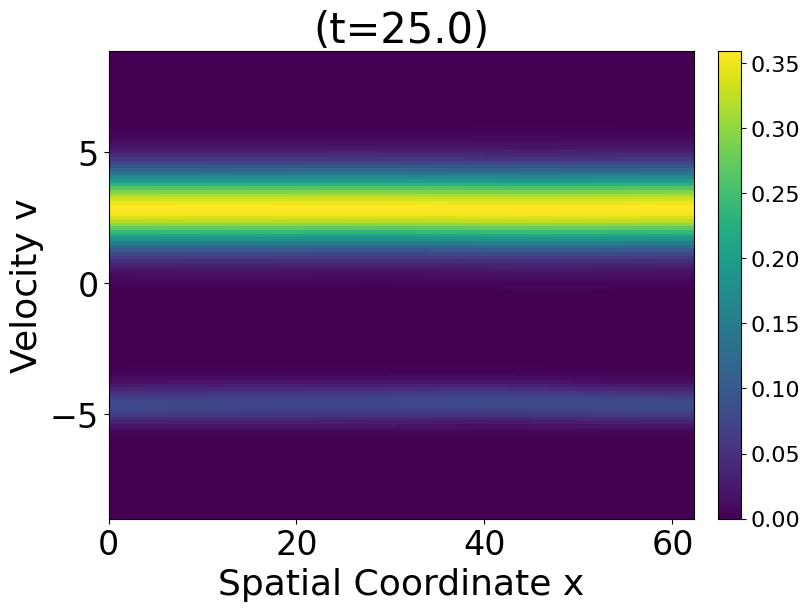}
        
        \vspace{0.5cm} 

        \includegraphics[width=0.25\textwidth]{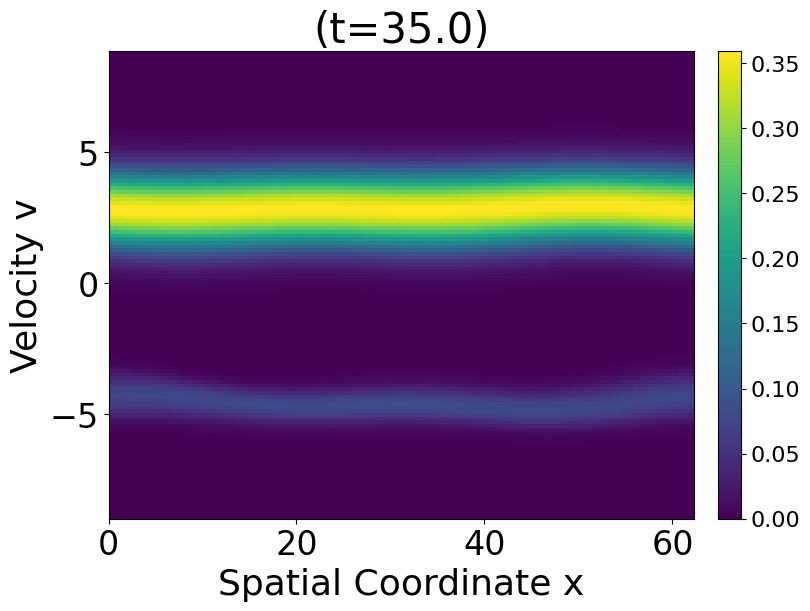}
        \includegraphics[width=0.25\textwidth]{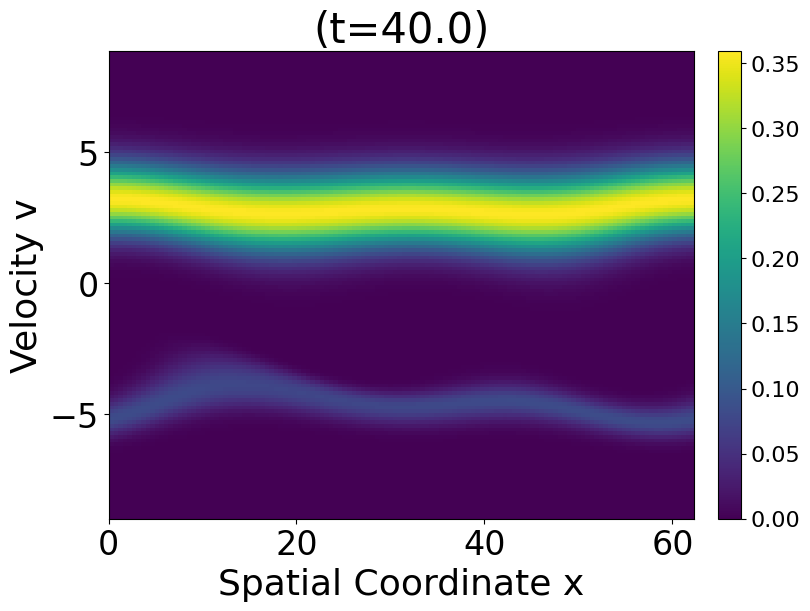}
        \includegraphics[width=0.25\textwidth]{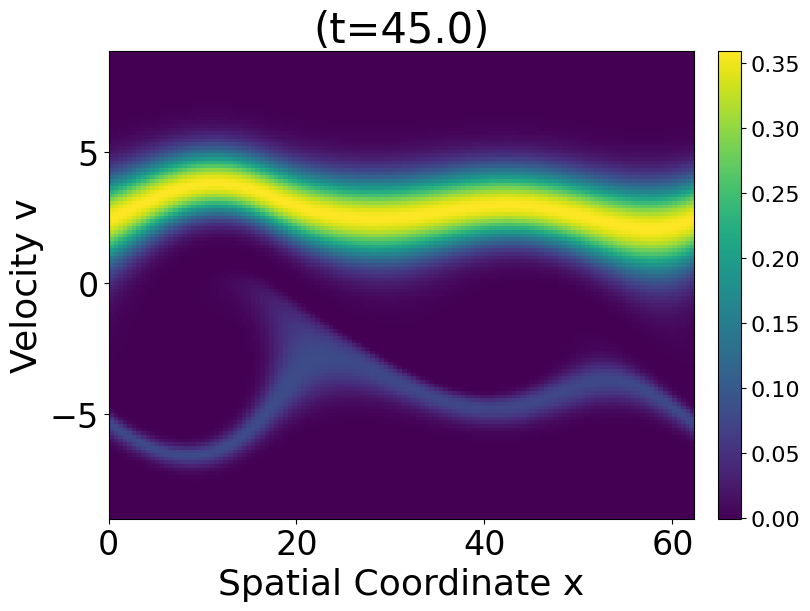}
        
        \vspace{0.5cm} 
        
        \includegraphics[width=0.25\textwidth]{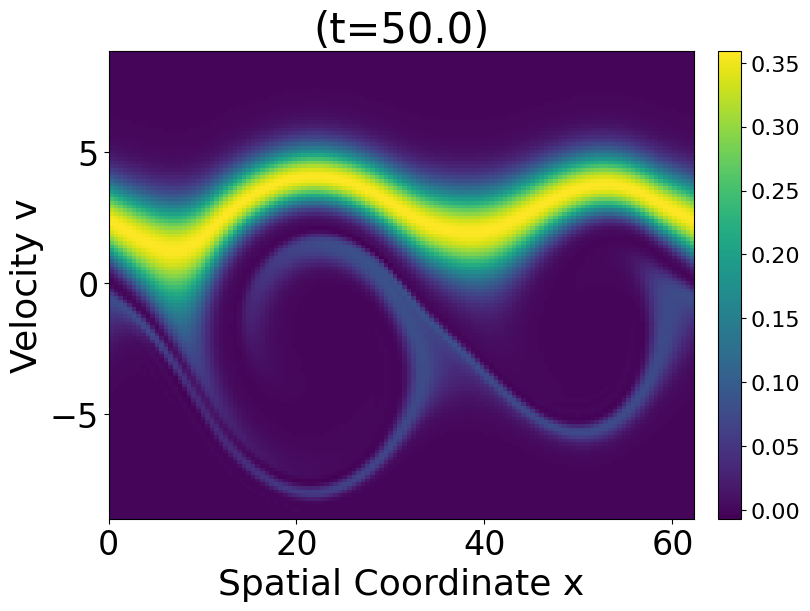}
        \includegraphics[width=0.25\textwidth]{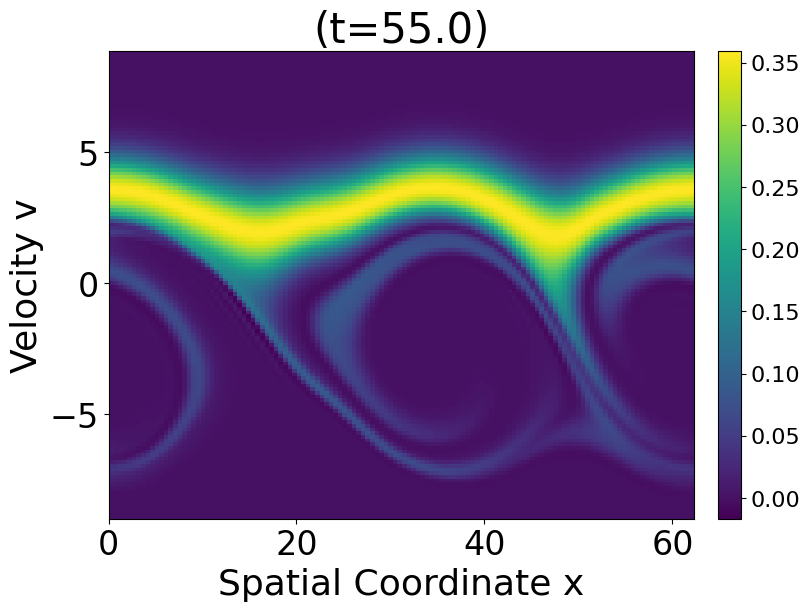}
        \includegraphics[width=0.25\textwidth]{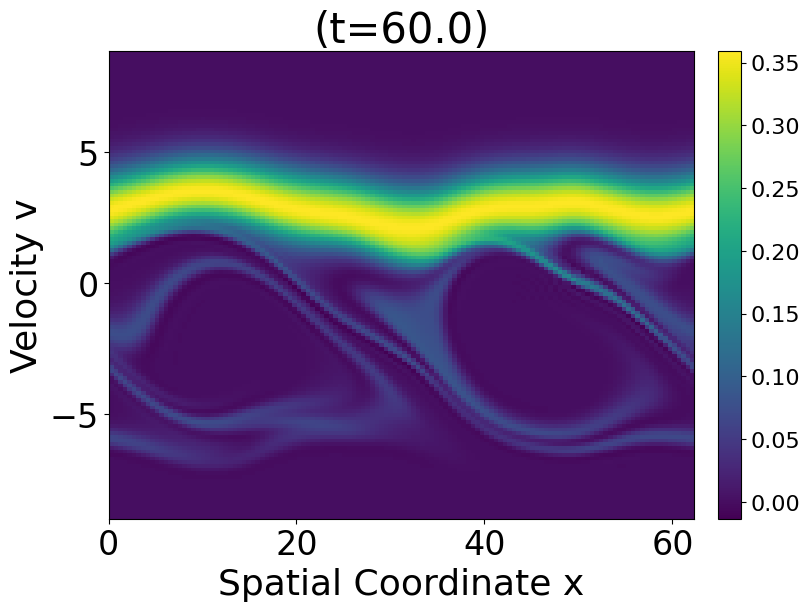}
        
        \caption*{(b) Snapshots for System C with $h(x) = -0.01 x$}
    \end{minipage}
    
    \caption{Comparative snapshots under different conditions for bump-on-tail case.}
    \label{fig:comparative_snapshots_bot}
\end{figure}

As a comparison to Figure \ref{fig:comparative_snapshots}, we illustrate the delayed process by comparing System A and System C in Figure \ref{fig:comparative_snapshots_bot}. We observe that without control in System A, instability appears as early as \(t=5\), starting from the tail, with a noticeable waving effect in the main Gaussian part around \(t=15\). In contrast, System C remains stable for a longer period. The obvious instability in System C begins between \(t=30\) and \(t=35\), which aligns with our observations of electric energy in Figure \ref{fig:electric_energy_comparative_bot}.

\begin{figure}[htb]
    \centering
    \includegraphics[width=0.46\textwidth]{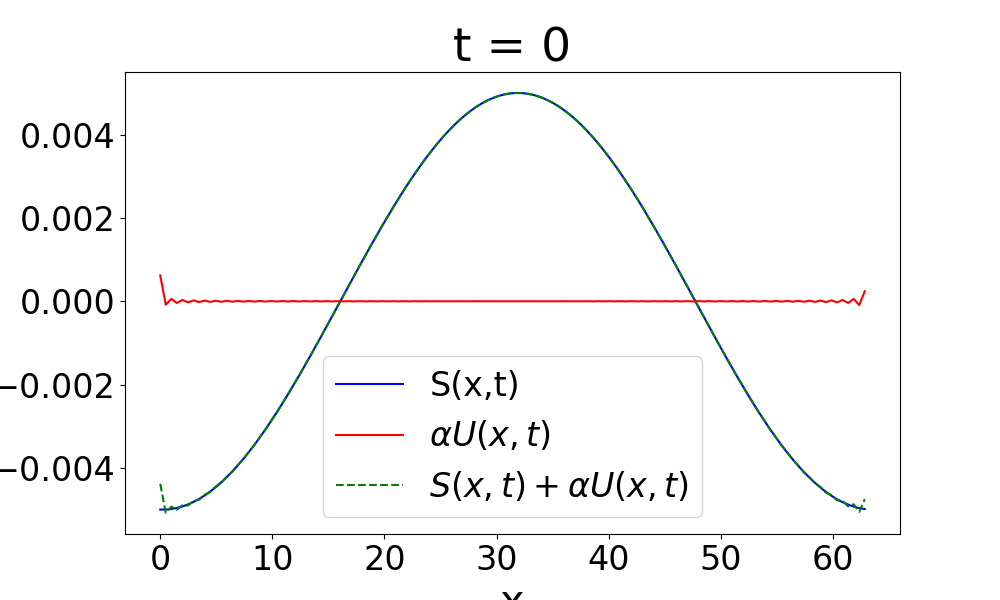}
    \includegraphics[width=0.48\textwidth]{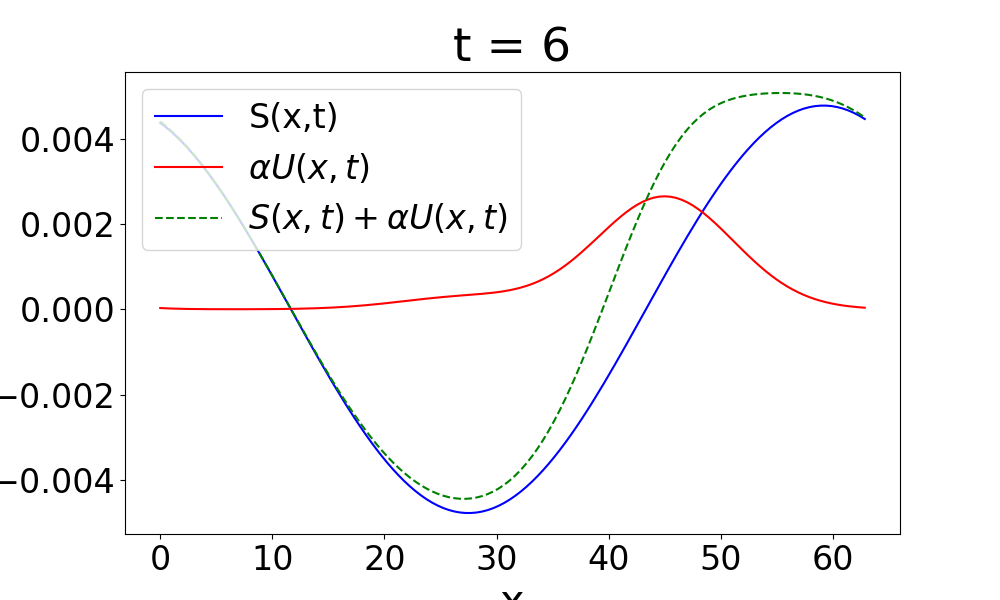}


     \vspace{0.5cm} 
    
    \includegraphics[width=0.48\textwidth]{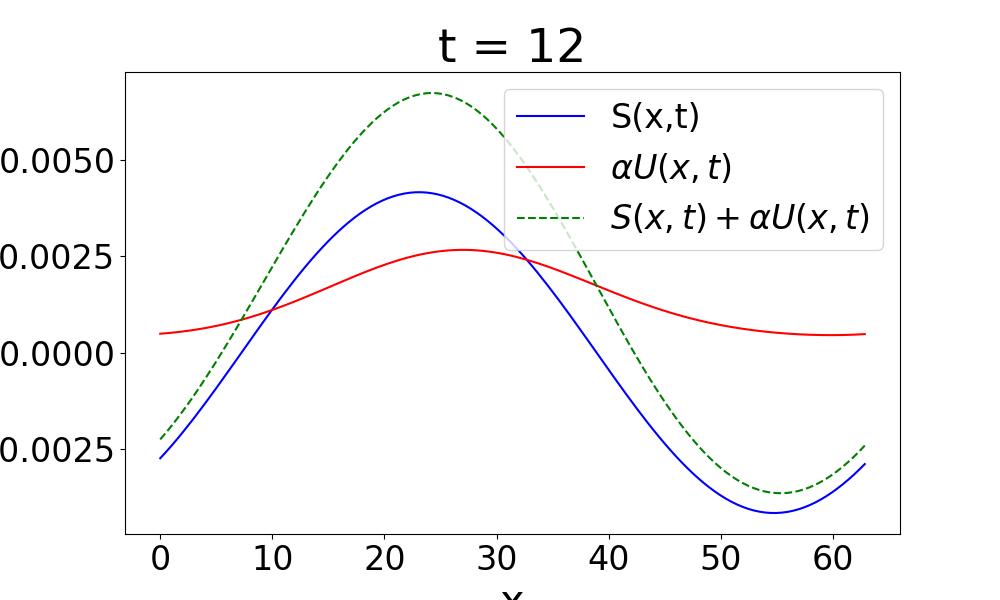}
    \includegraphics[width=0.48\textwidth]{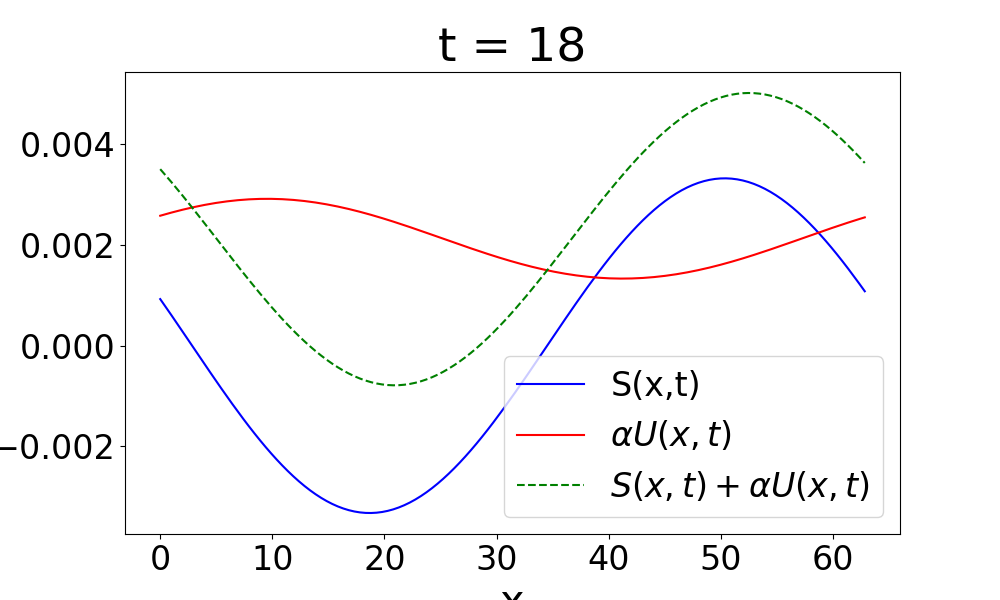}
    
    \caption{Graph of $S(t,x), \alpha U(t,x)$ and $S(t,x)+\alpha U(t,x)$ for bump-on-tail case at different moments}
    \label{fig:S+alphaU_bot}
\end{figure}

For numerical investigation, we further pursue the study of \( S \) and \( U \), similar to what have been done for the two-stream instability case. In Figure \ref{fig:S+alphaU_bot} we illustrate the evolution of \( \alpha U \), \( S \), and their combination. Initially, \( \alpha U \) exhibits a bump-on-tail structure, with the larger and smaller bumps gradually moving towards each other and merging. In contrast, \( S \) displays a sinusoidal pattern, propagating to a fixed direction. In time, there are moments when \( S \) and \( \alpha U \) effectively counter-balance each other, leading to a flattened combined curve, but there are also instances where they are in phase to each other, amplifying the combined effect. This makes the bump-on-tail instability more intricate than the two-stream instability.

\begin{figure}[htb!]
    \centering
    \includegraphics[width=5in]{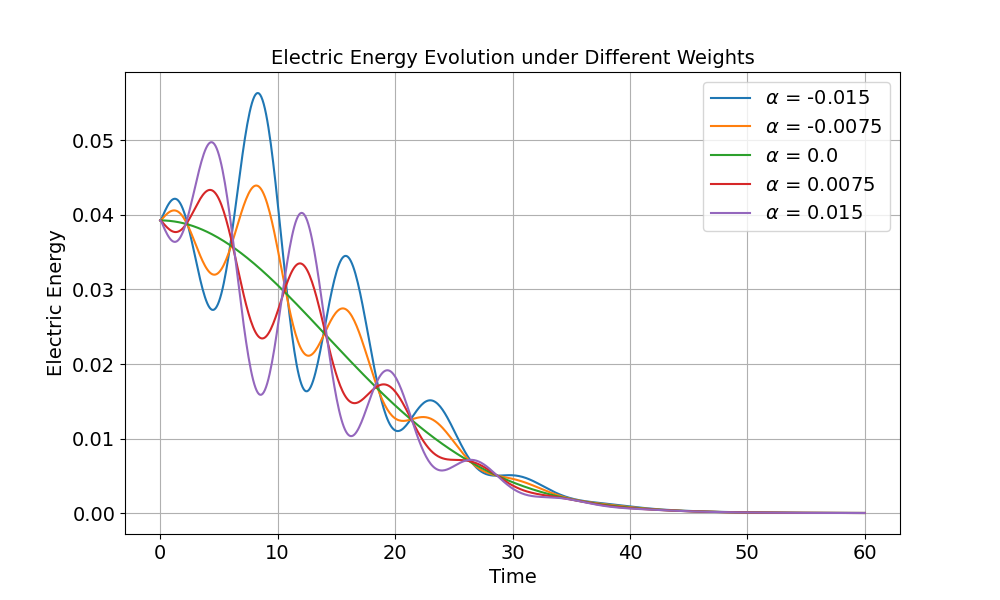}
    \caption{{Electric energy evolution for different weights used in System C when applied to the bump-on-tail instability}}
    \label{fig:electric_field_bot_weights}
\end{figure}

As a simple test for the effect of the weight \(\alpha\), we plot the electric energy for different values of \(\alpha \in [-0.015, -0.0075, 0, 0.0075, 0.015]\) in Figure \ref{fig:electric_field_bot_weights}. Recall that setting \(\alpha = 0\) recovers System B and corresponds to the free-streaming situation. It is evident that many choices of \(\alpha\) allow for better suppression of the electric energy than simple free-streaming for certain time phases. However, from this, we observe that a simple constant value for \(\alpha\) is unlikely to always be more optimal than free-streaming. A more complex structure might be needed for choosing \(\alpha\), and we leave this for future research.

\section{Conclusion}

In this study, we have focused on suppressing instabilities in plasma systems modeled by the Vlasov-Poisson (VP) equation, by introducing an external electric field. Through the application of Fourier-Laplace analysis to the linearized VP system, we derived an explicit formulation that illustrates the dynamic interplay between the evolution of the density function and the applied external field. Building upon this formula, we introduced a comprehensive framework that outlines an analytical strategy for the precise design of external fields. This methodology makes the external field explicitly computable once the initial perturbations within the plasma system and the desired equilibrium state are known. In a special scenario that we term electric field neutralization, this method revert the system to the free-streaming solution.

\section*{Acknowledgment}
\subsection*{Declaration of generative AI and AI-assisted technologies in the writing process}
During the preparation of this work, the author(s) used ChatGPT in order to improve the writing of this paper. After using this tool, the author(s) reviewed and edited the content as needed and take(s) full responsibility for the content of the publication.

\bibliographystyle{abbrv}
\bibliography{reference}

\appendix

\section{Proof of Proposition \ref{prop:drift_exponential_decay}}
\label{appen:proof_drift_decay}

\begin{proof}
   Let $f_\mathrm{f}(t,x,v)$ solve the free-streaming problem \eqref{eq:drift_equation_mod}, then $f_\mathrm{f}(t,x,v)$ has an explicit solution 
   \begin{equation}
       f_\mathrm{f}(t,x,v) = f_0(x-vt, v).
   \end{equation}
In Fourier space, this is equivalent to 
\begin{equation}
    \hat{f}_\mathrm{f}(t,k,\eta) = \hat{f}_0(k, \eta+kt) = \hat{\mu}(\eta+kt)\hat{X}(k),
\end{equation}
where $k$ and $\eta$ are Fourier variables corresponding to $x$ and $v$, and we have utilized the assumption that $f_0(x,v) = \mu(v)X(x)$. Using \eqref{eqn:s_interpretation}, we can see that
\begin{equation*}
    \hat{\rho}_\mathrm{f}(t,k) = \hat{f}_0(k,kt) = \hat{\mu}(kt)\hat{X}(k).
\end{equation*}
Using the fact that $\hat{X}(0)=0$, we can see that $\hat{\rho}_\mathrm{f}(t,0)=0$. And as the Fourier transform of a Gaussian is still Gaussian, $\hat{\rho}_\mathrm{f}(t,k)$ exhibits super-exponential decay in terms of $t$. Finally, the Plancherel theorem completes the proof.
   
\end{proof}

 \section{Proof of Proposition \ref{prop:external_Laplace}}
 \label{appen:proof_external_Laplace}

 \begin{proof}

To start, we introduce the function \( g(t, x, v) = f(t, x + vt, v) \) to facilitate our analysis. As a result, we have
$$\partial_t g(t,x,v)=\partial_t f(t,x+vt,v)+v\partial_xf(t,x+vt,v).$$
It transforms the first equation in the linear system \eqref{eq:VP_system_pert_linear} to be
\begin{equation}
\label{eq:vlasov_external_g}
    \partial_tg(t,x,v) - \left[E(t, x+vt)+H(t,x+vt)\right]\partial_v\mu(v)=0
\end{equation}
The density function $\rho(t,x)$ can be computed as
\begin{equation*}
\label{eq:rho_external}
    \rho(t,x) = \int_\mathbb{R} f(t,x,v)\,dv=\int_\mathbb{R}g(t,x-vt,v)\,dv.
\end{equation*}
As it has been mentioned before, here $\rho$ simply denotes the deviation of density from the equilibrium. Applying Fourier transform to $\rho$ and we obtain 
\begin{equation}
\label{eq:rho_external_Fourier}
    \begin{aligned}
           \hat{\rho}(t,k)&=\int_\mathbb{R} \rho(t,x) e^{-ikx}\,dx\\
           &=\int_\mathbb{R}\int_{\mathbb{R}}g(t,x-vt,v)e^{-ikx}\,dx\,dv\\
           &=\int_\mathbb{R}\int_{\mathbb{R}}g(t,x-vt,v)e^{-ik(x-vt)}\,e^{-ikvt}\,dx\,dv\\
           &=\hat{g}(t,k,kt),
    \end{aligned}
\end{equation}
where \( \hat{g} \) represents the Fourier transform of \( g \) with respect to both the spatial variable \( x \) and the velocity variable \( v \). Upon applying the Fourier transform to Equation \eqref{eq:vlasov_external_g}, we arrive at the following expression:

\begin{equation*}
    \begin{aligned}
0      &=\partial_t\hat{g}(t,k,\eta) - \int_\mathbb{R}\int_{\mathbb{R}}\left[E(t, x+vt)+H(t,x+vt)\right]e^{-ik(x+vt)}\partial_v\mu(v)e^{-i\eta v+ikvt}\,dx\,dv\\
&=\partial_t\hat{g}(t,k,\eta) - \left[\hat{E}(t,k)+\hat{H}(t,k)\right]\,\widehat{\partial_v\mu}(\eta-kt)\\
&=\partial_t\hat{g}(t,k,\eta) - i\left[\hat{E}(t,k)+\hat{H}(t,k)\right]\,(\eta-kt)\hat{\mu}(\eta-kt).
    \end{aligned}
\end{equation*}
Applying Fourier transform the electric field $E$, we get
\begin{equation}
\label{eq:E_formula_Fourier}
    ik\hat{E}(t,k)=-\hat{\rho}(t,k)
\end{equation}
and so we can link $\hat{H}$ and $\hat{E}$ by the following relation:
\begin{equation*}
    \partial_t\hat{g}(t,k,\eta) - i\hat{H}(t,k)\,(\eta-kt)\hat{\mu}(\eta-kt)+\frac{\hat{\rho}(t,k)}{k}(\eta-kt)\hat{\mu}(\eta-k t)=0
\end{equation*}
Integrating on both sides, we yield
\begin{equation*}
    \label{eq:external_g_Fourier}
\hat{g}(t,k,\eta)-\hat{g}(0,k,\eta)-i
\int_0^t\hat{H}(s,k)\,(\eta-ks)\hat{\mu}(\eta-ks)\,ds+\int_0^t\frac{\hat{\rho}(s,k)}{k}(\eta-ks)\hat{\mu}(\eta-ks)\,ds=0.
\end{equation*}
By setting \(\eta = kt\) and incorporating equation \eqref{eq:rho_external_Fourier}, we can further simplify and reframe this relation as
\begin{equation*}
    \label{eq:rho_Fourier_expression}
    \hat{\rho}(t,k)-ik\int_0^t\hat{H}(s,k)\,(t-s)\hat{\mu}\left[k(t-s)\right]\,ds +\int_0^t \hat{\rho}(s,k)(t-s)\hat{\mu}\left[k(t-s)\right]\,ds=\hat{g}(0,k,kt)=\hat{f}(0,k,kt),
\end{equation*}
where \( \hat{f} \) signifies the Fourier transform of \( f \). Both $\int_0^t\hat{H}(s,k)\,(t-s)\hat{\mu}\left[k(t-s)\right]\,ds$ and $\int_0^t\hat{\rho}(s,k)\,(t-s)\hat{\mu}\left[k(t-s)\right]\,ds$ present a convolution structure. Since the Laplace transform maps convolution to products we arrive at the following expression:
\begin{equation*}
    L[\hat{\rho}(\cdot,k)](s)-ik L[\hat{H}(\cdot,k)](s)\,L[\hat{U}(\cdot,k)](s)+L[\hat{\rho}(\cdot,k)](s)\,L[\hat{U}(\cdot,k)](s)=L[\hat{S}(\cdot,k)](s).
\end{equation*}
This is equivalent to the identity \eqref{eq:external_Laplace} that we aim to prove.
\end{proof}

\end{document}